\documentclass[11pt]{article}

\usepackage{amsmath,dsfont}
\usepackage[english]{babel}
\usepackage{amsfonts}
\usepackage{amsmath,amssymb}
\usepackage{graphicx,xcolor}
\usepackage{mathtools}

\makeatletter

\def\Z{\mathbb Z}

\def\J{\mathbb J}

\def\eqnarray{\stepcounter{equation}\let\@currentlabel=\theequation
\global\@eqnswtrue
\global\@eqcnt\z@\tabskip\@centering\let\\=\@eqncr
$$\halign to \displaywidth\bgroup\@eqnsel\hskip\@centering
  $\displaystyle\tabskip\z@{##}$&\global\@eqcnt\@ne
  \hfil$\displaystyle{\hbox{}##\hbox{}}$\hfil
  &\global\@eqcnt\tw@ $\displaystyle\tabskip\z@
  {##}$\hfil\tabskip\@centering&\llap{##}\tabskip\z@\cr}
\@addtoreset{equation}{section}
  \def\theequation{\thesection.\arabic{equation}}
\makeatother

\usepackage{amsmath,amssymb}
\def\beq{\begin{equation}}
\def\eeq{\end{equation}}
\def\beqa{\begin{eqnarray}}
\def\eeqa{\end{eqnarray}}
\def\barray{\begin{array}}
\def\earray{\end{array}}

\begin{document}

\title{{\bf    Particle in a self-dual dyon background:
hidden free nature, and exotic
 superconformal symmetry
}
}

\author{\textsf{
Mikhail S. Plyushchay${}^{a}$ and Andreas Wipf${}^{b}$ }
\\
[4pt]
 {\small \textit{${}^{a}$ Departamento de F\'{\i}sica, Universidad de
Santiago de Chile, Casilla 307, Santiago 2,
Chile}}\\
{\small \textit{${}^{b}$  Theoretisch-Physikalisches Institut,
Friedrich-Schiller-Universit\"at Jena,
Max-Wien-Platz 1,}}\\
{\textit{\small D-07743 Jena,
Germany }}\\
\sl{\small{E-mails:
mikhail.plyushchay@usach.cl, wipf@tpi.uni-jena.de} }}
\date{}

\maketitle

\begin{abstract}
We show that  a non-relativistic
particle in a combined field of a magnetic
monopole and  $1/r^2$ potential
reveals a hidden, partially
free dynamics when 
the strength of the central
potential and the charge-monopole 
coupling constant are mutually  fitted to each other. 
 In this case the system admits both
 a conserved Laplace-Runge-Lenz 
 vector and a dynamical
 conformal symmetry. 
The supersymmetrically extended
system corresponds then to 
a background of a self-dual or anti-self-dual dyon.
It is described by a quadratically  extended 
Lie  superalgebra  $D(2,1;\alpha)$  with 
$\alpha=1/2$,
in which the bosonic set of generators is enlarged  
by a generalized Laplace-Runge-Lenz vector and 
its dynamical integral counterpart related 
to Galilei symmetry, as well as by the chiral 
$\Z_2$-grading operator.  The odd part of the nonlinear 
superalgebra comprises a complete set of  
$24=2\times 3\times 4$ fermionic generators.
Here a usual duplication comes from 
the $\Z_2$-grading structure,
the second factor  can be 
associated with a triad of scalar integrals ---
the Hamiltonian, the generator 
of special conformal transformations
and the squared total angular momentum vector, while 
the quadruplication is generated by a chiral spin vector integral  
which exits due to the (anti)-self-dual nature
of the electromagnetic background.
\end{abstract}

\vskip.5cm\noindent

\section{Introduction}

Peculiar features  of  a  classical or quantum system are 
usually associated with and reflected in 
its special  symmetry properties.  A well-known example is 
the  conserved Laplace-Runge-Lenz vector,
which explains the periodicity of the classical bound
trajectories in the Kepler 
problem and 
the `accidental' degeneracy of the bound states energy 
levels of the hydrogen atom \cite{Pauli}--\cite{Wipf:2005se}.
A different   kind of example  is provided by
nonlinear integrable
systems, in which soliton solutions exhibit
particle-like  properties
in classical scattering processes. 
The robustness  of solitons in these field systems 
is a consequence of the infinite number of conservation laws.  
In the inverse scattering method, solitons
correspond to reflectionless  potentials 
in the associated quantum problems \cite{NovZak}. 
The reflectionless nature of soliton potentials can be linked,
in turn,  with a presence   of a nontrivial 
Lax-Novikov quantum  integral  of motion which is a higher 
order differential operator.
These  peculiarities of 
the  quantum mechanical soliton systems 
show up  in 
a supersymmetric generalization,  where 
they reveal a richer supersymmetry structure in 
comparison with that for the non-solitonic ones 
\cite{AMP}.

A charged  particle in the field of a magnetic monopole  exhibits
a hidden  free conical dynamics \cite{GoddOl,PlMon}.
In this aspect
it resembles
one-dimensional quantum mechanical reflectionless
systems with their  close relation to a free particle.
A charge-dyon system,  on the other hand,  
is characterized by the presence
of the conserved Laplace-Runge-Lenz vector  \cite{Zwanz},
similarly to the Kepler problem.
The study of both the charge-monopole and charge-dyon 
systems, as well as their superextensions, has
attracted a lot of attention in literature \cite{GoddOl}--\cite{Hong}. 

 This paper is devoted to the investigation of the rather exotic 
 nonlinear
superconformal structure of a particle 
in the field of the Dirac magnetic
monopole accompanied by the field of the central 
$1/r^2$-potential. 
Particular aspects of this system, including the supersymmetric one, have
been investigated in earlier works \cite{DHokVin3,HorP,Ivanov+}. 
In the present work we shall, however, emphasize the aspects
related to the hidden symmetries.
Namely, we first investigate in detail the spinless particle
and show, that for a particular value of the strength of
the central potential relative to the charge-monopole 
coupling, the system
reveals a hidden partially free dynamics. 
As a result,
besides the rotational and conformal symmetries, 
it will admit
the conserved Laplace-Runge-Lenz vector as 
well as the associated dynamical (\emph{explicitly} 
depending on time)
 vector integral
related to the Galilei symmetry.   Then we shall 
arrive at  a related system from a 
different direction,
by constructing the supersymmetric extension of the 
particle in an electromagnetic background field
 by incorporating  spin degrees of freedom. 
 We shall observe that for a (anti)-self-dual background
 the system admits a chiral spin integral of motion.
  As a result, we obtain a supersymmetric generalization 
  of the original
 spinless system,  which can be treated as a 
 supersymmetric spinning particle 
 in the field of a (anti)-self-dual dyon.
 The supersymmetric structure we obtain is rather
 unusual and unexpectedly rich.
 It incorporates the 
 Laplace-Runge-Lenz and  the 
 associated dynamical vector
 integrals,  the generators of  conformal 
 and rotational symmetries,  and the 
 chiral spin vector integral. 
 They enter the resulting partially nonlinear 
 (quadratic) superalgebra 
 with 24 quantum fermionic generators,
 which represents a certain extension of the 
 superconformal $D(2,1;\alpha)$ symmetry 
 \cite{Ivanov+} with 
 a particular value of  the parameter $\alpha=1/2$.

 In the following  section we investigate the  
spinless particle  and in particular the special case 
 characterized by
 the presence of the conserved 
 Laplace-Runge-Lenz vector  and a partially 
 free dynamics. 
 In the third section we  construct the 
 supersymmetric extension of the system, and study
 its nonlinear superconformal structure 
 both at the classical and quantum 
 levels.
 The last, fourth section includes a summary and  concluding remarks.

\section{Spinless case}

Consider a non-relativistic particle of charge $e$ and mass $m$ in
a combined field of a magnetic 
monopole\footnote{We use the units $c=\hbar=1$.}, 
$\vec{B}=g\vec{r}/r^3$, and
central potential $U(r)$.
It is described by the Hamiltonian 
\begin{equation}\label{H}
    H=\frac{1}{2m}\vec{\Pi}^2+U(r),
\end{equation}
and Poisson brackets
\begin{equation}\label{PBs}
    \{r_i,r_j\}=0,\qquad
    \{r_i,\Pi_j\}=\delta_{ij},\qquad
    \{\Pi_i,\Pi_j\}=e\epsilon_{ijk}B_k\,.
\end{equation}
The  equations of motion for the position 
vector and kinetic-momentum read
\begin{equation}\label{EqMo}
    \dot{\vec{r}}=\frac{1}{m}\vec{\Pi},\qquad
    \dot{\vec{\Pi}}=-\frac{\nu}{mr^3}\vec{L}-U'(r)\vec{n}\,,
\end{equation}
where $\nu=eg$,  $\vec{n}=\vec{r}/r$,
$\vec{L}=\vec{r}\times\vec{\Pi}$. {}From here one finds  
\begin{equation}\label{nL}
    \dot{\vec{n}}=\frac{1}{mr^2}\vec{L}\times\vec{n}\,,\qquad
    \dot{\vec{L}}=\frac{\nu}{mr^2}\vec{L}\times\vec{n}\,,
\end{equation}
and
\begin{equation}\label{rPi}
    \frac{d}{dt}\vec{r}{}\,^2=\frac{2}{m}\vec{\Pi}\cdot\vec{r}\,,\qquad
    \frac{d}{dt}(\vec{\Pi}\cdot \vec{r})=2H-(2U+rU')\,.
\end{equation}
{}From (\ref{nL}) it follows that 
the Poincar\'e vector
\begin{equation}\label{J}
    \vec{J}=\vec{L}-\nu\vec{n}\,,\quad
    \text{with}\quad
    \vec{J}^{\,2}=\vec{L}^{\,2}+\nu^2\geq \nu^2\,,
\end{equation}
is an integral of motion for any choice of 
the central potential.
It is just the angular momentum
of the system\,:
\begin{equation}\label{JJn}
    \{J_i,J_j\}=\epsilon_{ijk}J_k\,,\qquad
    \{J_i,n_j\}=\epsilon_{ijk}n_k\,,\qquad
    \{J_i,r\}=\{J_i,{\Pi}_r\}=0\,,
\end{equation}
where $\Pi_r=\vec{\Pi}\cdot \vec{n}$ is the radial component
of the kinetic momentum and we also have
$
    \{r,\Pi_r\}=1.
$
In terms of the 
variables $\vec{J}$, $\vec{n}$,
$r$ and $\Pi_r$
 the Hamiltonian takes the form
\begin{equation}\label{H}
    H=\frac{1}{2m}\left(
    {\Pi}_r^2 +\frac{(\vec{J}\times\vec{n})^2}{r^2}\right) +U(r),\qquad
    \text{with}\qquad
    (\vec{J}\times\vec{n})^2=\vec{J}^{\,2}-\nu^2\,.
\end{equation}
The vectors $\vec{n}$ and 
$\vec{L}$ precess around  the conserved angular momentum
 $\vec{J}$ with the same frequency,
 \begin{equation}\label{nLJ}
    \dot{\vec{n}}=\frac{1}{mr^2}\vec{J}\times\vec{n}\,,\qquad
    \dot{\vec{L}}=\frac{1}{mr^2}\vec{J}\times\vec{L}\,.
\end{equation}
Hence the trajectory of the particle lies on the cone defined by
$\vec{J}\cdot\vec{n}=-\nu$ 
with vertex in $r=0$ and symmetry axis
oriented along the vector $\vec{J}$.
For $U(r)=0$
the particle moves on geodesics on the cone \cite{GoddOl,PlMon},
and like a free particle ($\nu=0$) is characterized 
by a conformal symmetry \cite{conf1}-\cite{confla+}. 
This symmetry survives 
under switching on the inverse square potential 
\begin{equation}\label{Ur2}
    U(r)=\frac{\lambda}{r^2}\,.
\end{equation}
In this case the scalar $\vec{\Pi}\cdot\vec{r}$ is subject to 
a simple dynamics,
$\frac{d}{dt}(\vec{\Pi}\cdot\vec{r})=2H$.  As a consequence,
the
dilatation generator
\begin{equation}\label{D}
    D=\vec{\Pi}\cdot\vec{r}-2tH
\end{equation}
is an explicitly time-dependent dynamical 
integral of motion:
$\frac{d}{dt}D=\frac{\partial}{\partial t}D+\{D,H\}=0$. The first
equation from (\ref{rPi}) implies then that 
another dynamical integral of motion exists:
\begin{equation}\label{K}
    K=2mr^2-4tD-4t^2H\,.
\end{equation}
It is the generator of special conformal transformations.

 {}From
now on, we shall consider the potential
(\ref{Ur2}) characterized by the presence of the two dynamical
integrals of motion $D$ and $K$, and assume that $\lambda> 0$ to
avoid the problem of the fall to the center  $r=0$. 
We shall see that the system admits an even 
richer symmetry structure when
the relation $\lambda=\nu^2/2m$ between 
the couplings holds true. This particular choice of couplings is
also distinguished by the supersymmetric extension of the system.

The minimal distance of the particle from the 
force center corresponds to the instant
 $t_0$ for which
$(\vec{\Pi}\cdot\vec{r})(t_0)=0$, 
see Eq. (\ref{rPi}). Taking into
account relation 
$\vec{L}^2=2mr^2H- (\vec{\Pi}\cdot\vec{r})^2-\mu^2$,
where $\mu^2=2m\lambda$, one finds that
$
    r_{\rm min}^2=(\vec{J}^{\,2}-\nu^2+\mu^2)
    /(2mH).
$
Now we decompose the unit vector
$\vec{n}$  into the parts
 $\vec{n}_{\vert\vert}$ and $\vec{n}_{\perp}$
parallel and perpendicular to 
the total angular momentum.
Due to (\ref{J}) the former is time independent,
$\vec{n}_{\vert\vert}=-\nu\vec{J}/\vec{J}{\,}^2$,
and this implies $\vec{n}_{\perp}^2=(\vec{J}^{\,2}-\nu^2)/
\vec{J}^{\,2}$. Thus the latter can be written as 
\begin{equation}\label{n_perp}
    \vec{n}_{\perp}(t)=\vec{n}_{\perp}(t_0)\cos\varphi(t)+
    \frac{1}{J}\vec{J}\times\vec{n}_{\perp}(t_0)\sin\varphi(t).
\end{equation}
Since
$\dot{\vec{n}}_{\perp}=\dot{\vec{n}}$  the
first relation in
(\ref{nLJ}) implies
$\dot{\varphi}=J/(mr^2)$.
Employing the relations
$1/r^2=2mH/[(\vec{\Pi}\cdot\vec{r})^2+\vec{L}^{\,2}+\mu^2]$ and
$2H=\frac{d}{dt}(\vec{\Pi}\cdot\vec{r})$, we obtain $d\varphi=
Jd(\vec{\Pi}\cdot\vec{r})/[(\vec{\Pi}\cdot\vec{r})^2+\vec{L}^{\,2}+\mu^2]$.
This yields the evolution law for the angle $\varphi$ in the plane
orthogonal to the angular momentum, 
\begin{equation}\label{phit}
    \varphi(t)=\frac{J}{\sqrt{\vec{L}^{\,2}+\mu^2}}
    \arctan\left(\vec{\Pi}\cdot\vec{r}/{\sqrt{\vec{L}^2+\mu^2}}\right)
    +\hbox{const.}
\end{equation}
For $t_0=0$ and a vanishing integration constant the
angle vanishes when $r$ is minimal, 
$
    \vec{\Pi}\cdot \vec{r}=2Ht,
 $
 and
 $
    r^2(t)=r^2_{\rm min}+2Ht^2/m\,,
$
such that
\begin{equation}\label{phit}
    \varphi(t)=\frac{J}{\sqrt{J^2-\nu^2+\mu^2}}
    \arctan\left(\frac{2Ht}{\sqrt{J^2-
    \nu^2+\mu^2}}\right)\,,
\end{equation}
\begin{equation}\label{nnn}
    \vec{n}(t)=-\nu \frac{1}{J^2} \vec{J}+\vec{n}_{\perp}(t)\,,\qquad
     \vec{n}_{\perp}(t)=(\vec{n}(0)+\nu \frac{1}{J^2}\vec{J}\,)\cos\varphi(t)+
    \frac{1}{J} \vec{J}\times\vec{n}(0)\sin\varphi(t)\,.
\end{equation}
The scattering angle of
the trajectory projected onto the plane
orthogonal to $\vec{J}$ is
\begin{equation}\label{delphi}
    \Delta\varphi-\pi=\int_{-\infty}^{+\infty}\dot{\varphi}dt -\pi=
    \pi\left(\frac{J}{
    \sqrt{J^2-\nu^2+\mu^2}}-1\right)\,.
\end{equation}
In general it depends on the value of the angular momentum 
$J$. Only in the exceptional case when  $\mu^2=\nu^2$
we have $\Delta\varphi-\pi=0$ for all  $J^2\geq \nu^2$.

From now on we restrict our analysis to the special case
$\mu^2=\nu^2$, for which the Hamiltonian can be presented in
the two equivalent forms
\begin{equation}\label{Hspec}
    H=\frac{1}{2m}\left(\vec{\Pi}^2+
    \frac{\nu^2}{r^2}\right)=
    \frac{1}{2m}
    \left(\Pi_r^2+\frac{\vec{J}{\,}^2}{r^2}\right)\,.
\end{equation}
Only for this particular value of the parameter
$\lambda$
the central potential compensates exactly 
the term $-\nu^2/r^2$ appearing in the 
centrifugal  term $(\vec{J}\times\vec{n})^2/r^2$ 
of the charge-monopole Hamiltonian, see Eq. (\ref{H}).

The  cos- and sin-functions, which enter the precession law
of the unit vector, simplify (only)  in this case to elementary
functions, $\cos\varphi(t)={1}/{\sqrt{1+\tau^2}}$,
$\sin\varphi(t)=\tau/{\sqrt{1+\tau^2}}$, where we introduced the
dimensionless time variable
$
\tau={2Ht}/{J}
$.
The distance of the particle from the force center varies as
$ r(t)={J}\sqrt{1+\tau^2}/{\sqrt{2mH}},$
and we get
\begin{equation}\label{rminspec}
    r_{\rm min}=r(0)=\frac{J}{\sqrt{2mH}}\,.
\end{equation}
The evolution law for the particle's
coordinate vector can then be
written as
\begin{equation}\label{vecrt}
    \vec{r}(t)=\vec{r}(0)+\frac{J}{\sqrt{2mH}}\left(
    \frac{\nu}{J^2}\,\vec{J}\,\left(1-\sqrt{1+\tau^2}\right)+
   \frac{1}{J} \left(\vec{J}\times\vec{n}(0)\right)\,\tau\right).
\end{equation}
The projected motion of the particle
in the plane orthogonal to the angular 
momentum is that of a \emph{free particle}:
 it moves along a straight line with constant velocity.

Peculiar properties of a dynamical 
system both at the classical and quantum levels are in
many cases associated with the 
presence of hidden symmetries. 
This happens also in the special
case of the system (\ref{Hspec}). 
Indeed, consider the 
Laplace-Runge-Lenz-vector
\begin{equation}\label{GvecU}
    \vec{G}=\vec{\Pi}
    \times\vec{J}+\kappa\vec{n}\,,
\end{equation}
where $\kappa$  is a constant with the dimension of a mass.
In the general case with central potential
(\ref{H}) its dynamics  is
given by 
\begin{equation}\label{Kdyn}
    \frac{d}{dt}{\vec{G}}=
    \vec{L}\times\vec{n}\left(U'+\frac{\nu^2}{mr^3}+
    \frac{\kappa}{mr^2}\right).
\end{equation}
The vector is an integral of motion 
only for central potentials of the form
\begin{equation}\label{Uspec}
    U(r)=\frac{\nu^2}{2mr^2}+\frac{\kappa}{mr}\,,
\end{equation}
that is for a linear combination of 
the Kepler potential and the particular potential
 (\ref{Ur2}) with
\begin{equation}\label{lga}
    2m\lambda=\mu^2=\nu^2\,.
\end{equation}
 The case $\kappa=0$ is characterized
by the presence of the additional conformal symmetry associated
with dynamical integrals $D$ and $K$. 
The particular system
(\ref{Hspec}) 
therefore admits the additional
integral of motion
\begin{equation}\label{Gvec0}
    \vec{G}=\vec{\Pi}\times\vec{J}
    \,.
\end{equation}
This, particularly, can easily be seen from 
the equation of motion
\begin{equation}\label{pidot}
    \dot{\vec{\Pi}}=-\frac{\nu}{mr^3}\vec{J}
\end{equation}
which holds for relation
 (\ref{lga}) between the couplings, see 
 Eqs. (\ref{EqMo}) and
(\ref{J}).

{}From relations (\ref{Gvec0}) and (\ref{J}) one also finds
\begin{equation}\label{Gr}
    \vec{G}\cdot\vec{r}=J^2-\nu^2.
\end{equation}
Eq. (\ref{vecrt}) 
in particular means, that the vector
$\vec{J}\times \vec{n}(0)$ is oriented along the integral
\begin{equation}\label{AJ}
    \vec{G}\times
    \vec{J}=-J^2\vec{\Pi}-\nu\Pi_r\vec{J}\,.
\end{equation}
The conserved vectors
 $\vec{J}$, $\vec{G}$ and $\vec{G}\times \vec{J}$ form an
orthogonal basis, and in addition to (\ref{Gr}), 
the
projections of
 $\vec{r}$ and $\vec{\Pi}$ onto these
vectors are
\begin{equation}\label{rpro}
    \vec{r}\cdot\vec{J}=-\nu r\,,\qquad
    \vec{r}\cdot(\vec{G}\times\vec{J})=-r\Pi_r(J^2-\nu^2)\,,
\end{equation}
\begin{equation}\label{pipro}
    \vec{\Pi}\cdot\vec{G}=0\,,\qquad
    \vec{\Pi}\cdot(\vec{G}\times
    \vec{J})=-\vec{G}^{\,2}\,,\qquad
    \vec{\Pi}\cdot\vec{J}=-\nu\Pi_r\,.
\end{equation}
In addition, note that
\begin{equation}\label{G2}
    \vec{G}^{\,2}=
    2mH({{J}}^2-\nu^2)\,,\qquad
    (\vec{G}\times\vec{J})^2=\vec{G}^{\,2} J^2\,.
\end{equation}
The angle between the 
vectors $\vec{n}(t)$ and $\vec{J}$ is given by
$\cos\theta=-\nu J^{-1}$. Taking into account  Eq.
(\ref{rminspec}), one finds that relation (\ref{Gr}) can be
written in the equivalent form
\begin{equation}\label{plane}
    \vec{G}\cdot\left(\vec{r}(t)-\vec{r}(0)\right)=0\,.
\end{equation}
The trajectory of the particle lies, therefore, in the plane
orthogonal to  $\vec{G}$.  
We conclude that the
trajectory is given by intersection of the cone
$\vec{J}\cdot \vec{n}=-\nu$
with the specified plane.
It has a form of a
hyperbola, whose projection onto the plane orthogonal to $\vec{J}$
is a straight line parallel to the conserved vector $\vec{G}\times
\vec{J}$. The projected coordinate of the particle evolves with 
constant speed along this line.  
The equation of hyperbola can be
presented in the form
\begin{equation}\label{hyper}
    \frac{(\vec{r}\cdot\vec{J}\,)^2}{\nu^2 J^2}-
    \frac{(\vec{r}\cdot(\vec{G}\times
    \vec{J}))^2}{G^2 J^2(J^2-\nu^2)}
    =\frac{1}{2mH}\,.
\end{equation}
Since the conserved vectors
 $\vec{J}$,
$\vec{G}$ and  $\vec{G}\times \vec{J}$ form the complete
orthogonal set in the $3$-dimensional space, one finds
\begin{equation}\label{r(t)}
    \vec{r}=\frac{1}{2mH}\,\vec{G}-
    \frac{\nu r}{J^2}\,\vec{J}-
    \frac{\vec{r}\cdot\vec{\Pi}}{2mH J^2}\,
    \vec{G}\times \vec{J}\,,
\end{equation}
where the relations
\begin{equation}\label{rtev}
    (\vec{r}\cdot\vec{\Pi})(t)=
    (\vec{r}\cdot\vec{\Pi})(0)+2Ht\,,\qquad
    r^2(t)=r^2(0)+\frac{2}{m}(\vec{r}\cdot\vec{\Pi})(0)t
    +\frac{2}{m}Ht^2
\end{equation}
finally determine the evolution of $\vec{r}(t)$. One obtains 
the same law as in
 (\ref{vecrt}). 

In addition to the scalar integrals $D$ and $K$, explicitly
depending on time, the first equation in (\ref{EqMo}) and Eq.
(\ref{Gvec0}) allow us 
to construct an analogous, 
dynamical vector integral depending
explicitly on time,
\begin{equation}\label{Rint}
    \vec{{R}}=\vec{r}\times\vec{J}-
    \frac{t}{m}\,
    \vec{G}\,.
\end{equation}
It satisfies the relations
\begin{equation}\label{RNJ}
    2mH\vec{R}=D\vec{G}+\vec{G}\times\vec{J}\,,
\end{equation}
and we also get
\begin{equation}\label{RJA}
    \vec{{R}}\cdot\vec{J}=0\,,\qquad
    \vec{{R}}\cdot(\vec{G}\times
    \vec{J})=J^2(J^2-\nu^2)\,,
\end{equation}
\begin{equation}\label{RJADK}
    \vec{{R}}\cdot\vec{G}=(J^2-\nu^2)D\,,\qquad
    \vec{{R}\,}^2=\frac{1}{2m}(J^2-\nu^2)K\,.
\end{equation}

In the liming case $\nu\rightarrow 0$
corresponding to a free particle, one gets
\begin{equation}\label{Rlimal}
     \vec{{R}}\rightarrow
     D\left(\vec{r}-\frac{t}{m}\vec{p}\right)
     -\frac{K}{2m}\vec{p}\,.
\end{equation}
In more detail,  at $g=\nu=0$,   the mechanical (or kinetic) momentum
$\vec{\Pi}$  turns into the canonical 
momentum  $\vec{p}$ with
Poisson-commuting components, $\{p_i,p_j\}=0$.
At the same time the angular momentum $\vec{J}$
becomes the orbital angular momentum 
and the system transforms into a free particle with
$H=\vec{p}^{\,2}/2m$. In accordance with Eq. (\ref{AJ}),
the integral $\vec{G}\times \vec{J}$ reduces to canonical momentum
vector $\vec{p}$ multiplied by the integral $-\vec{L}^{2}$.
The Laplace-Runge-Lenz vector $\vec{G}$ 
itself reduces to $\vec{p}\times
\vec{L}=\vec{p}^{\,2}\vec{r}-(\vec{p}\cdot\vec{r})\vec{p}$.
Note that free particle system 
possesses the additional dynamical integral
 $\vec{N}=\vec{r}-\vec{p}\,t/m$
 that coincides with $\vec{r}(0)$, and is 
 a generator of Galilei boosts.
It is interesting to 
compare the free particle relations  
$\vec{N}\cdot\vec{p}=D$ and $2m\vec{N}^2=K$
with (\ref{RJADK}).
The integral
$\vec{p}\times \vec{L}$ can be written in terms of the dynamical
integrals $\vec{N}$ and $D$ as $\vec{p}\times
\vec{L}=2mH\vec{N}-D\vec{p}$, which can be compared  
with the limit relation (\ref{RNJ}) for
the dynamical vector integral $\vec{R}$. 
The trajectory of the free particle
is a straight line along the vector $\vec{p}$ that passes though
the point $\vec{r}(0)=\vec{N}$ and lies in the plane orthogonal to
$\vec{L}$. 
 Switching
on the magnetic monopole field and at the same time
the scalar potential $U(r)=\nu^2/2mr^2$ 
results in  `lifting' and deforming the straight
line into the hyperbola given by 
the intersection of the magnetic
monopole cone $\vec{J}\cdot\vec{n}=-\nu$
with a plane orthogonal to
$\vec{G}$ and passing through
 the point $\vec{r}(0)$.

{}From (\ref{G2}) it follows that the Hamiltonian of the
system can
be presented in terms of the angular momentum and
Laplace-Runge-Lenz vector,
\begin{equation}\label{HGJ}
    H=\frac{1}{2m}
    \frac{{{G}}^2}{{{J}}^2-\nu^2}\,.
\end{equation}
The latter satisfies the Poisson bracket relation
\begin{equation}\label{GG}
    \{{G}_i,{G}_j\}=
    -2mH\epsilon_{ijk}J_k\,.
\end{equation}
As $H>0$,  one  defines the vector
\begin{equation}\label{VGH}
    \vec{V}=\frac{\vec{G}}{\sqrt{2mH}}\,.
\end{equation}
This re-scaled Laplace-Runge-Lenz vector together with $\vec{J}$
generate the $so(3,1)$ Lorentz algebra,
\begin{equation}\label{JSJ}
    \{J_i,J_j\}=\epsilon_{ijk}J_k\,,\qquad
    \{{V}_i,{V}_j\}=
    -\epsilon_{ijk}J_k\,,\qquad
    \{J_i,{V}_j\}=\epsilon_{ijk}{V}_k\,.
\end{equation}
The quantities ${\mathcal{C}}_1=\vec{V}^{\,2}-\vec{J}^{\,2}$ and
${\mathcal{C}}_2=\vec{J}\cdot\vec{V}$ are 
two independent Casimirs of the $so(3,1)$ algebra (\ref{JSJ}),
$\{{\mathcal{C}}_a, J_i\}=\{{\mathcal{C}}_a, {V}_i\}=0$, $a=1,2$,
which have here the values ${\mathcal{C}}_1=\nu^2$
and 
${\mathcal{C}}_2=0$. In terms of the complex combinations
$
    {\mathcal{L}}_j^\pm=\frac{1}{2}(J_j\pm i{V}_j)\,,
$
we have
\begin{equation}\label{LL+-}
    \{{\mathcal{L}}_i^+,{\mathcal{L}}_j^+\}=
    \epsilon_{ijk}{\mathcal{L}}_k^+\,,\qquad
    \{{\mathcal{L}}_i^-,{\mathcal{L}}_j^-\}=
    \epsilon_{ijk}{\mathcal{L}}_k^-\,,\qquad
    \{{\mathcal{L}}_i^+,{\mathcal{L}}_j^-\}=0\,.
\end{equation}

In conclusion of this section, let us note that we have 
identified additional integrals of motion
for particular central potential by first 
analyzing the scattering of the particle.
The acceleration points in the direction of $\vec{J}$
only if  $U(r)=\nu^2/(2mr^2)+\hbox{const}$, i.e.
exactly for the particular central 
potential we have studied.
For this potential
$\vec{G}=\vec{\Pi}\times \vec{J}$ is an 
integral of motion. 
The acceleration of $\vec{r}$, projected on 
the integral $\vec{J}\times \vec{G}$, is zero
and this reveals a hidden partially 
free dynamics of the particle. The relation
$\frac{d^3}{dt^3}(r^2)=0$ is equivalent to the 
condition that $K$ is a \emph{dynamical} integral of motion.
Since $\vec{r}\cdot\vec{J}=-\nu r$,  this reduces to
the equation
$\frac{d^3}{dt^3}(\vec{r}\cdot\vec{J}\,)^2=0$. The last relation means 
that the acceleration of the particle along
$\vec{J}$ is constant,
and from here we recover the hyperbolic form of the trajectory.

Note that  the system (\ref{Hspec}) corresponds to
a spinless part  of the model  \cite{DHokVin3} 
at the ``points of higher symmetry" $\lambda^2=\nu^2$.
It was discussed in \cite{Ivanov+}, where  
a special hyperbolic trajectory was also identified and 
associated  with the presence of the Laplace-Runge-Lenz vector. 
However, there the dynamical integral (\ref{Rint})
was not considered. As we shall see below, 
both the vector integrals $\vec{G}$ and $\vec{R}$ 
(more precisely, their analogs incorporating  
spin degrees of freedom) will
play the key role in a  
nonlinear supeconformal structure of the 
superextended version of the system.

In the next section we find a supersymmetric extension of the system 
 (\ref{Hspec})  by exploiting its
particular symmetry properties.
 
\section{Supersymmetric extension:  particle 
in (anti)-self-dual dyon background}

To construct a supersymmetric generalization of the system,
we introduce four Grassmann variables $\xi_a$, 
where $a=0,i$,  and $i=1,2,3$, 
with Poisson brackets
\begin{equation}\label{xixi}
    \{\xi_a,\xi_b\}=-i\delta_{ab}\,.
\end{equation}
Their quantum analogs  are 
given by Euclidean gamma-matrices $\gamma_a$, 
$\hat{\xi}_0=\frac{1}{\sqrt{2}}\gamma_0$,
 $\hat{\xi}_i=\frac{1}{\sqrt{2}}\gamma_i$,
 realized, for example, via 
 two sets of the Pauli matrices,
\begin{equation}
	\gamma_0=\tau_1\otimes 1=
\left(
\begin{array}{ccc}
 {\bf 0} &    {\bf 1}   \\
 {\bf 1}  &    {\bf 0}   \\   
\end{array}
\right),\quad
	\gamma_i=\tau_2\otimes \sigma_i=
\left(
\begin{array}{ccc}
 {\bf 0} &   -i \sigma_i   \\
 i\sigma_i  &    {\bf 0}   \\   
\end{array}
\right).
\end{equation}
We distinguish the values $a=0$ and $a=1,2,3$
since in the model we shall obtain  the $\xi_0$ and $\xi_i$ 
will have different transformation properties  
under the spatial rotations.   
The operators $\hat{\xi}_a$ anti-commute with 
\begin{equation}\label{Gamma}
\Gamma\equiv\gamma_5=
 \tau_3\otimes 1=\left(
\begin{array}{cc}
 {\bf 1} &    {\bf 0}   \\
  {\bf 0}  &    -{\bf 1}   \\   
\end{array}
\right),\qquad
\Gamma^2=1\,,
\end{equation}
which at the quantum level 
is identified 
as a $Z_2$-grading operator
of the superalgebraic structure.
We introduce also the chiral projectors
\begin{equation}\label{TTT}
\mathcal{T}_\pm=\frac{1}{2}(1\pm\gamma_5)=
\frac{1}{2}(1\pm\tau_3)\otimes 1\,,\qquad
\mathcal{T}_++\mathcal{T}_-=1\,,\qquad
\mathcal{T}_+\mathcal{T}_-=0\,.
\end{equation}
In terms of the Grassmann variables 
$\xi_i$ and $\xi_0$, one defines 
the chiral spin vectors:
\begin{equation}
    \mathcal{S}^\pm_i=\frac{1}{2}
    (\mathcal{S}_i\pm \mathcal{V}_i)\,,\quad \text{where}\quad
    \mathcal{S}_i=-\frac{i}{2}\epsilon_{ijk}\xi_j\xi_k\,,\quad
    \mathcal{V}_i=-i\xi_0\xi_i\,.
\end{equation}
They generate the $so(4)=so(3)\oplus so(3)$ algebra,
\begin{equation}\label{LLL}
    \{\mathcal{S}^+_i,\mathcal{S}^+_j\}=\epsilon_{ijk}
    \mathcal{S}^+_k\,,\quad
    \{\mathcal{S}^-_i,\mathcal{S}^-_j\}=\epsilon_{ijk}
    \mathcal{S}^-_k\,,\quad
    \{\mathcal{S}^+_i,\mathcal{S}^-_j\}=0\,.
\end{equation}
The $\mathcal{S}^+_i$  has the following Poisson 
brackets with the $\xi_a$\,:
\begin{equation}\label{L+xi}
    \{\mathcal{S}^+_i,\xi_j\}=\frac{1}{2}(\epsilon_{ijk}\xi_k
    -\xi_0\delta_{ij})\,,\qquad
    \{\mathcal{S}^+_i,\xi_0\}=\frac{1}{2}\xi_i\,.
\end{equation}
Analogous relations  for $\mathcal{S}^-_i$ are obtained from
(\ref{L+xi}) by the change $\xi_0\rightarrow -\xi_0$.

The quantum analogs of $ \mathcal{S}^\pm_i$ contain the 
chiral projectors, $ \hat{\mathcal{S}}^\pm_i=\mathcal{T}_\pm\hat{S}_i$,
where $\hat{S}_i=\hat{\mathcal{S}}^+_i+\hat{\mathcal{S}}^-_i=
1\otimes \frac{1}{2}\sigma_i $, and, particularly,  
the quantum analog
of the third relation from  (\ref{LLL}),
$[\hat{\mathcal{S}}^+_i,
\hat{\mathcal{S}}^-_j]=0$,
 is just a trivial consequence of the
opposite chiralities  of $\hat{\mathcal{S}}^+_i$ and 
$\hat{\mathcal{S}}^-_i$.

Now we consider a particle with charge 
$e$ 
propagating in an electric and magnetic fields
described by a vector potential $A_i(\vec{r})$ 
and a scalar potential $A_0(\vec{r})$,
and  consider a Grassmann-odd classical quantity
\begin{equation}\label{Theta0phi}
    \Theta_0=\Pi_i\xi_i+\phi(\vec{r})\xi_0\,,
\end{equation}
where $\phi(\vec{r})=eA_0(\vec{r})$ and
$\Pi_i(\vec{r})=p_i-eA_i(\vec{r})$
are the components of the kinetic momentum
$\vec{\Pi}$. 
{}From here on we 
set $2m=1$. We have
$\{\Pi_i,\Pi_j\}=\epsilon_{ijk}\mathcal{B}_k$
and $\{\Pi_i,\phi\}=\mathcal{E}_i$ 
with $\mathcal{B}_i=eB_i$ and  $\mathcal{E}_i=eE_i$.
Here $B_i=\epsilon_{ijk}\partial_jA_k$ and $E_i=-\partial_iA_0$
are the  background magnetic and electric fields,
whose forms are not further specified at the moment.

The
Grassmann-even quantity $\mathcal{H}$ generated
by $\Theta_0$,
\begin{equation}\label{QQHdef}
    \{\Theta_0,\Theta_0\}=-i\mathcal{H}\,,
\end{equation}
\begin{equation}\label{Hspingen}
    \mathcal{H}=\Pi_i^2+\phi^2-2
    \left(\mathcal{S}^+_i(\mathcal{B}_i-
    \mathcal{E}_i)+
    \mathcal{S}^-_i(\mathcal{B}_i+
    \mathcal{E}_i)\right)\,,
\end{equation}
is readily identified as a Pauli type, second order  
in $p_i$  Hamiltonian. 
With such an interpretation, $\Theta_0$ can be considered 
as a classical
analog of the stationary ($\partial/\partial t\rightarrow 0$),
 first order in $p_i$ Dirac operator.
{}From the generalized, graded Jacobi
identity $3\{\Theta_0,\{\Theta_0,\Theta_0\}\}=0$
it follows at once that $\{\Theta_0,\mathcal{H}\}=0$, 
and so, $\Theta_0$ can
be treated as a supercharge for the system 
with the Hamiltonian $\mathcal{H}$.
 Eq. (\ref{Hspingen}) shows that
  independently from the rotational properties of potentials
 $A_i$ and $A_0$, there are  
 two  special cases\footnote{We are not interested here
in  another special case corresponding to homogeneous electric
and magnetic fields.}: self-dual,
when $\mathcal{B}_i=\mathcal{E}_i$, and anti-self-dual,
$\mathcal{B}_i=- \mathcal{E}_i$. As follows from the 
last relation in
(\ref{LLL}), in these two  cases we have additional Grassmann-even
integrals of motion, $\mathcal{S}^+_i$, or $\mathcal{S}^-_i$,
respectively. Since with any anti-self-dual  background one can
associate the corresponding self-dual background just by changing
$A_0\rightarrow -A_0$,  $A_i\rightarrow A_i$,
one can restrict the consideration to the self-dual
case. Then the Hamiltonian reduces
to
\begin{equation}\label{Hsefldyon}
\mathcal{H}=\Pi_i^2+\phi^2-4 \mathcal{S}^-_i\mathcal{B}_i\,,
\end{equation}
and in addition to the supercharge $\Theta_0$
we have the integrals of motion 
$\mathcal{S}^+_i$ 
generating an
$so(3)$ symmetry. 
As the Poisson brackets of 
integrals of motion are also integrals of motion,  
 we get three more integrals
\begin{equation}
    \Theta_i\equiv \epsilon_{ijk}\Pi_j\xi_k
    +\phi\xi_i-\Pi_i\xi_0\,,
\end{equation}
where $\Theta_i=2\{\mathcal{S}^+_i,\Theta_0\}$.
Let us stress that the integrals
$\Theta_i$ and $\Theta_0$ form the set with the same
transformation properties with respect to the $so(3)$ generators
$\mathcal{S}^+_i$  as the set formed by 
the basic Grassmann variables $\xi_i$ and
$\xi_0$,
\begin{equation}\label{L+Qi}
    \{\mathcal{S}^+_i,\Theta_0\}=
    \frac{1}{2}\Theta_i\,,\qquad
    \{\mathcal{S}^+_i,\Theta_j\}=\frac{1}{2}
    \left(\epsilon_{ijk}
    \Theta_k-\delta_{ij}\Theta_0\right)\,.
\end{equation}
Employing (\ref{L+Qi}), together with the conservation of
$\mathcal{S}^+_i$ and the graded Jacobi identity 
$-\{\Theta_0,\{\mathcal{S}^+_i,
\Theta_0\}\}+ \{\mathcal{S}^+_i,\{\Theta_0,
\Theta_0\}\} +\{\Theta_0,\{
\Theta_0,\mathcal{S}^+_i\}\}=0$, 
we find that $\Theta_i$ and $\Theta_0$
Poisson-commute,
\begin{equation}\label{Th0Thi}
    \{\Theta_0,\Theta_i\}=0\,,
\end{equation}
similarly to
$\xi_i$ and $\xi_0$.
Once again using the Jacobi identity and relations (\ref{L+Qi}) and
(\ref{QQHdef}), we get
\begin{equation}\label{ThiThj}
    \{\Theta_i,\Theta_j\}=-i\delta_{ij}\mathcal{H}\,.
\end{equation}

Thus, postulating the supercharge (\ref{Theta0phi}) and choosing 
the (anti)self-dual electromagnetic background, we have got 
the second order in momenta $p_i$ system possessing  
the $so(3)\cong su(2)$ symmetry, whose 
generators   $\mathcal{S}^+_i$ give rise 
to the extension of  the $N=1$ supersymmetry 
(\ref{QQHdef}) up to the $N=4$ supersymmetry 
(\ref{QQHdef}), (\ref{Th0Thi}), (\ref{ThiThj}). 
Note here that for the first time it was showed  
in  \cite{CromRit} that 
the $N=4$ supersymmetry for a  particle in three-dimensional
space 
necessarily implies the self-duality of the 
electromagnetic background, see also 
\cite{DHokVin3,Ivanov+,IvaLech}.

Up to this point
we did not assume  any
particular properties of the background field
with respect to the spatial rotations.
Suppose now that electric field is spherically symmetric 
and choose $\phi=\phi(r)$. Then
$\vec{\mathcal{E}}=-{\vec{r}}\phi'(r)/{r}=
\vec{\mathcal{B}}$, and  the  Maxwell equation
$\partial_i\mathcal{B}_i=0$ for $\vec{r}\neq 0$
fixes the magnetic field to be that of the magnetic
monopole, $\vec{\mathcal{B}}=\nu{\vec{r}}/{r^3}$.
We arrive therefore at the electromagnetic background
of the self-dual dyon characterized in general case by 
the scalar potential $\phi(r)=\kappa+\nu/r$,
where $\kappa$ is a constant. 
Hamiltonian (\ref{Hsefldyon}) is then a supersymmetric generalization
of the spinless case given by 
 the potential (\ref{Uspec}). 
 We are interested 
 in the supersymmetric generalization
 of the special case characterized by a hidden partially free 
 particle dynamics. So, we  put $\kappa=0$, 
and the Hamiltonian of the
system takes the form (\ref{Hsefldyon}) with
$\phi={\nu}/{r}$,
\begin{equation}\label{Hsefldyon+}
\mathcal{H}=\Pi_i^2+\frac{\nu^2}{r^2}-4\nu\frac{1}{r^3}
 \mathcal{S}^-_i{r}_i\,.
\end{equation}
Its Grassmann-free, spinless  part 
coincides with the Hamiltonian (\ref{Hspec}), which 
possesses
the dynamical conformal symmetry,
and one can expect that the total symmetry  of 
(\ref{Hsefldyon+}) has to be a supersymmetric extension
of that of  the spinless system (\ref{Hspec}).
Having in mind this perspective, we shall show
now how an exotic nonlinear (quadratic) 
superconformal algebra for
 (\ref{Hsefldyon+}) appears by exploiting our knowledge
about the symmetries of the system (\ref{Hspec}).

First, we find that the total angular momentum 
$\mathcal{J}_i=J_i+\mathcal{S}_i$
is an integral of motion of the system (\ref{Hsefldyon+}), where
$J_i=(\vec{r}\times\vec{\Pi})_i-\nu n_i$.
With
respect to it, the 
$\xi_i$ form a vector and $\xi_0$  a scalar, and the
Grassmann-odd 
 supercharges $\Theta_i$ and $\Theta_0$ have 
 exactly the same rotational 
properties.
The 
$\mathcal{S}^+_i$ is the Grassmann-even vector integral.
As a result we find that 
the total angular momentum $\mathcal{J}_i$ and 
the chiral spin $\mathcal{S}_i^+$ generate 
the bosonic $so(4)=so(3)\oplus so(3)$ symmetry,
\begin{equation}
	\{\mathcal{Y}_i^+,\mathcal{Y}_j^+\}=\epsilon_{ijk}
	\mathcal{Y}_k^+,\quad
	\{\mathcal{Y}_i^-,\mathcal{Y}_j^-\}=\epsilon_{ijk}
	\mathcal{Y}_k^-,\quad
	\{\mathcal{Y}_i^+,\mathcal{Y}_j^-\}=0\,,
\end{equation}
where
\begin{equation}
	\mathcal{Y}_i^+\equiv \mathcal{S}_i^+,\qquad
	\mathcal{Y}_i^-\equiv\mathcal{J}_i-\mathcal{S}_i^+\,.	
\end{equation}
The Poisson bracket relations for the supercharges $\Theta_a$ 
with the bosonic integrals $\mathcal{Y}_i^\pm$ are
\begin{equation}\label{JpmT}
	\{\mathcal{Y}_i^\pm,\Theta_0\}=\pm\frac{1}{2}
	\Theta_i,\qquad
	\{\mathcal{Y}_i^\pm,\Theta_j\}=\frac{1}{2}
	(\epsilon_{ijk}
	\Theta_k\mp\delta_{ij}\Theta_0)\,.
\end{equation}

As in the spinless case,
the superextended system also possesses dynamical integrals
corresponding to scale- and special conformal transformations.
The corresponding Grassmann-even dynamical scalar 
 integrals  are
\begin{equation}\label{DK}
    \mathcal{D}=\vec{\Pi}\cdot\vec{r}-2\mathcal{H}t\,,\qquad
    \mathcal{K}=\vec{r}^{\,2}-4\mathcal{D}t-4\mathcal{H}t^2\,.
\end{equation}
Indeed one finds 
$\frac{d}{dt}\mathcal{I}= \{\mathcal{I}, \mathcal{H}\}+
\frac{\partial}{\partial t}\mathcal{I}=0$, where 
$\mathcal{I}=\mathcal{D},\,\mathcal{K}$.
Together with the
Hamiltonian, they generate the $so(2,1)$ algebra,
\begin{equation}\label{so(2,1)}
    \{\mathcal{D},\mathcal{H}\}=2
    \mathcal{H}\,,\quad
    \{\mathcal{D},\mathcal{K}\}=-2
    \mathcal{K}\,,\quad
    \{\mathcal{K},\mathcal{H}\}=4
    \mathcal{D}\,.
\end{equation}
In terms of the standard basis
\begin{equation}\label{Josp}
	\J_0=\frac{1}{4}(\mathcal{H}
	+\mathcal{K})\,,\qquad
	 \J_1=\frac{1}{4}(\mathcal{H}
	-\mathcal{K})\,,\qquad
	 \J_2=\frac{1}{2}\mathcal{D}\,,
\end{equation} 
the $so(2,1)$-structure is manifest,
$\{\J_0,\J_1\}=\J_2$, $\{\J_0,\J_2\}=-\J_1$,
$\{\J_1,\J_2\}=-\J_0$.
All three generators  
Poisson-commute with the integrals $\mathcal{J}_i$ and
$\mathcal{S}^+_i$, i.e.
\begin{equation}\label{DKJS}
    \{\mathcal{Y}_i^\pm,\J_\mu\}=0\,,
    \qquad \mu=0,1,2\,.
\end{equation}
The Poisson brackets of $\mathcal{K}$ with $\Theta_0$ and
$\Theta_i$ generate the Grassmann-odd dynamical integrals,
\begin{equation}\label{Omega0i}
    \Omega_0=\rho_0-2\Theta_0t\,,\qquad
    \Omega_i=\rho_i-2\Theta_it\,,
\end{equation}
where 
$$
\rho_0=\vec{r}\cdot\vec{\xi}\,, \qquad
\vec{\rho}=\vec{r}\times\vec{\xi}-\xi_0\vec{r}.
$$
The Grassmann-odd quantities  $(\rho_0,\vec{\rho}\,)$ 
are transformed by the chiral spin integral $\mathcal{S}^+_i$ 
and total angular momentum $\mathcal{J}_i$ in the same 
way as $(\xi_0,\vec{\xi}\,)$.  
The  
$\Omega_0$ and $\Omega_i$ are, respectively, the scalar and vector
integrals with respect to $\mathcal{J}_i$,
$\{\mathcal{J}_i,\Omega_0\}=0$, $\{\mathcal{J}_i,\Omega_j\}=
\epsilon_{ijk}\Omega_k$, while with respect to the chiral spin
vector $\mathcal{S}^+_i$ they transform in the same way 
as the integrals
$\Theta_0$ and $\Theta_i$. Hence
 the dynamical integrals
$\Omega_a$ have the Poisson 
bracket relations of the form (\ref{JpmT})
with the $so(4)$ generators 
$\mathcal{Y}_i^\pm$.
The Poisson brackets of $\Theta_a$ and $\Omega_a$ 
with the $so(2,1)$ generators are
\begin{equation}\label{DTO}
    \{\mathcal{D},\Theta_a\}=\Theta_a\,,\qquad
    \{\mathcal{D},\Omega_a\}=-\Omega_a\,,
\end{equation}
\begin{equation}\label{HTO}
    \{\mathcal{H},\Theta_a\}=0\,,\qquad
    \{\mathcal{H},\Omega_a\}=-2\Theta_a\,,
\end{equation}
\begin{equation}\label{KTO}
    \{\mathcal{K},\Theta_a\}=2\Omega_a\,,\qquad
    \{\mathcal{K},\Omega_a\}=0\,.
\end{equation}
The Poisson bracket relations between the Grassmann-odd
integrals are 
\begin{equation}\label{ThOm}
    \{\Theta_a,\Theta_b\}=-i\delta_{ab}\mathcal{H}\,,\qquad
    \{\Omega_a,\Omega_b\}=-i\delta_{ab}\mathcal{K}\,,
\end{equation}
and
\begin{equation}\label{OmThe}
    \{\Theta_0,\Omega_0\}=-i\mathcal{D}\,,\qquad
     \{\Theta_i,\Omega_j\}=i\epsilon_{ijk}(\mathcal{J}_k
     -4\mathcal{S}^+_k)-i\delta_{ij}\mathcal{D}\,,
\end{equation}
\begin{equation}\label{Om0Thi}
    \{\Theta_0,\Omega_i\}=-i(\mathcal{J}_i
    +2\mathcal{S}^+_i)\,,\qquad
    \{\Theta_i,\Omega_0\}=i(\mathcal{J}_i
    +2\mathcal{S}^+_i)\,.
\end{equation}
Thus, the  set  $\mathcal{H}$, $\mathcal{J}_i$,
$\mathcal{S}^+_i$, $\mathcal{D}$, $\mathcal{K}$ of even, 
 and the  set 
$\Theta_a$, $\Omega_a$ of odd  integrals together form 
 a closed Lie superalgebra. To identify it,
we represent the Poisson bracket relations 
(\ref{JpmT}), (\ref{OmThe}) and  (\ref{Om0Thi}) 
in a compact form:
\begin{equation}
	\{\mathcal{Y}^\pm_i,\Upsilon_a\}=
	\frac{1}{2}t^{\pm i}_{ab}\,\Upsilon_b\,,\qquad
	\{\Theta_a,\Omega_b\}=-i\mathcal{D}
	\delta_{ab}+2i\left(\alpha\, t^{- i}_{ab}\mathcal{Y}^-_i
	-(1+\alpha)\,
	t^{+ i}_{ab}\mathcal{Y}^+_i
	\right),
\end{equation}
where $\Upsilon_a=\Theta_a$, $\Omega_a$,
\begin{equation}
	t^{\pm i}_{ab}=-t^{\pm i}_{ba}\,,\qquad
	t^{\pm i}_{0j}=\pm\delta^i_j\,,\qquad
	t^{\pm i}_{jk}=\epsilon_{ijk},
\end{equation}
and $\alpha=1/2$.
We conclude that 
the nine bosonic integrals 
$\mathcal{H}$, $\mathcal{K}$, $\mathcal{D}$, 
$\mathcal{Y}^+_i$ and $\mathcal{Y}^-_i$, and the eight
fermionic integrals $\Theta_a$ and $\Omega_a$ generate
the superconformal  $D(2,1;\alpha)$  
symmetry
 \cite{VDJ,FSS,BMSV,confla,IKL}\footnote{Superalgebra
$D(2,1;\alpha)$  has an automorphism
associated  with permutations 
of the three $so(3)$ subalgebras, which are 
generated
by $\mathcal{Y}^+_i$, $\mathcal{Y}^-_i$
and $\tilde{\mathcal{Y}}_i$, where $\tilde{\mathcal{Y}}_1=i\J_1$,
$\tilde{\mathcal{Y}}_2=i\J_2$, $\tilde{\mathcal{Y}}_3=\J_0$.
At  the parameter level the automoprphism corresponds 
to the dihedral group 
$D_3$ generated by the transformations 
$\alpha\rightarrow -(1+\alpha)$ 
and $\alpha\rightarrow
\alpha^{-1}$.  
As a result,  the superalgebras
$D(2,1;\lambda)$ with $\lambda=\alpha^{\pm1}$, 
$-(1+\alpha)^{\pm1}$ and
$-(\frac{\alpha}{1+\alpha})^{\pm 1}$ are 
isomorphic  \cite{FSS},
and the superalgebra 
we have here can be identified as the
$D(2,1;\alpha)$  with 
the parameter $\alpha$  taking any value 
from the set $\{-3,-3/2,-2/3,-1/3,1/2,2\}$.}  
with
$\alpha=1/2$.
The three bosonic integrals
$\mathcal{H}$, $\mathcal{K}$, $\mathcal{D}$ together with
a pair of fermionic integrals $\Theta_a$, $\Omega_a$ with fixed $a$ 
generate one of the four copies of the $osp(1\vert 2)$ Lie 
superalgebra. As a minimal generating set 
one can take, for example, the odd integrals  
$\Theta_0$ and $\Omega_0$, and 
the even  integrals   $\mathcal{S}^+_1$ and 
$\mathcal{S}^+_2$. 
Their successive Poisson brackets fully reproduce the described 
superconformal Lie superalgebra.
The quadratic  
Casimir element of the Lie superalgebra  $D(2,1;\alpha)$
can be presented in
 the form \cite{MatMor,FIOL}
\begin{equation}\label{Casimir}
	\mathcal{C}=\tilde{\mathcal{Y}}_i
	\tilde{\mathcal{Y}}_i 
	+\alpha\mathcal{Y}^-_i\mathcal{Y}^-_i
	-(1+\alpha)\mathcal{Y}^+_i\mathcal{Y}^+_i
	+
	\frac{i}{2}\Theta_a\Omega_a\,.
\end{equation}
In our case 
$\tilde{\mathcal{Y}}_1=i\J_1$,
$\tilde{\mathcal{Y}}_2=i\J_2$, $\tilde{\mathcal{Y}}_3=\J_0$,
$\tilde{\mathcal{Y}}_i \tilde{\mathcal{Y}}_i=\J_0^2-
	\J_1^2-\J_2^2=\frac{1}{4}(\mathcal{H}
	\mathcal{K}-\mathcal{D}^2)$,
and one can easily  check that 
at
$\alpha=1/2$
the $\mathcal{C}$ 
given by Eq. (\ref{Casimir})
Poisson commutes with all the even and odd generators
$\mathcal{H}$, $\mathcal{K}$, $\mathcal{D}$,
$\mathcal{Y}^\pm_i$ and $\Theta_a$, $\Omega_a$.

We have here
$\frac{i}{2}\Theta_a\Omega_a=
(\vec{L}+\nu\vec{n})\cdot\vec{\mathcal{S}}^-$, and 
the quantum analog of the last nilpotent term in
(\ref{Casimir})
is
$$
	\frac{i}{4}[\hat{\Theta}_a,\hat{\Omega}_a]=
	\hat{\mathcal{L}}_\sigma +
	\frac{3}{2}\,,\quad \text{where}\quad
	\hat{\mathcal{L}}_\sigma\equiv
	\mathcal{T}_-(1\otimes(\hat{\vec{L}}+
	\nu\vec{n})\cdot\vec{\sigma}).
$$
This  is a nontrivial integral
for the spin-1/2 subsystem with Hamiltonian
$\hat{H}_-=\mathcal{T}_-\hat{\mathcal{H}}$,
see below.
It satisfies the relation
$\hat{\mathcal{L}}_\sigma(
\hat{\mathcal{L}}_\sigma+2)
=\big(\hat{\vec{\mathcal{J}}}^2-3/4\big)
\mathcal{T}_-$.
A rather natural question at this point is whether the system possesses 
a fermionic type  integral  which (like the Grassmann-even integral
$i\Theta_a\Omega_a/2$) quantum mechanically 
would be the square root of the 
integral $\hat{\vec{\mathcal{J}}}^2$ 
(possibly shifted for an additive constant)
but without a chiral projector factor.
Now we shall show that such a Grassmann-odd integral 
indeed exists, and that it is associated 
with the  conserved  Laplace-Runge-Lenz vector
of the superextended system.

To that end consider the Grassmann-odd scalar quantity
\begin{equation}\label{Xi0}
    \Xi_0=-\left(\vec{L}+\frac{2}{3}\vec{\mathcal{S}}
    \right)\cdot\vec{\xi}\,.
\end{equation}
One can check that it satisfies the Poisson-bracket relation
\begin{equation}\label{Xi0Xi0}
    \{\Xi_0,\Xi_0\}=-i\left(\vec{\mathcal{J}}^{\,2}-
    \nu^2\right),
\end{equation}
and  that it is an integral of motion, $\{\Xi_0,\mathcal{H}\}=0$.
The Poisson bracket of (\ref{Xi0}) with the chiral spin
vector $\mathcal{S}^+_i$ generates 
then three more integrals of motion,
which form a Grassman vector with respect to
the total angular momentum,
\begin{equation}\label{Xii}
    \vec{\Xi}=\vec{\xi}\times\vec{L}+\xi_0
    (\vec{L}+2\vec{\mathcal{S}})\,.
\end{equation}
With respect to the chiral spin vector, the integrals  $\Xi_a$ 
have properties similar to those of 
$\Theta_a$ and $\Omega_a$,
\begin{equation}\label{SXi}
    \{\mathcal{S}^+_i,\Xi_0\}=\frac{1}{2}\Xi_i\,,\qquad
    \{\mathcal{S}^+_i,\Xi_j\}=\frac{1}{2}
    \left(\epsilon_{ijk}
    \Xi_k-\delta_{ij}\Xi_0\right)\,.
\end{equation}
There is, however, an essential difference in comparison with
the Grassmann-odd integrals $\Theta_a$ and $\Omega_a$. One can calculate 
the Poisson brackets of  $\Xi_0$ with $\Xi_i$ by 
using the first relation in (\ref{SXi}) 
and employing the graded Jacobi identities. 
Since the Poisson bracket of  $\mathcal{S}^+_i$  with 
the right hand side
in  (\ref{Xi0Xi0}) is nonzero, 
the integrals  $\Xi_0$ and $\Xi_i$
possess nontrivial Poisson bracket relations,
\begin{equation}\label{Xi0Xii}
    \{\Xi_0,{\Xi}_i\}=2i\,(\vec{\mathcal{S}}^+
    \times
    \vec{\mathcal{J}})_i\,.
\end{equation}
The presence of the quadratic in integrals 
$\vec{\mathcal{J}}$ and 
$\vec{\mathcal{S}}^+$
expressions on the right hand sides of (\ref{Xi0Xii})
and (\ref{Xi0Xi0})
means that the extension of the set of 
generators of the superconformal 
Lie algebra $D(2,1;\alpha=1/2)$ 
 by the odd integral $\Xi_0$ 
transforms it into a \emph{nonlinear}
superalgebra, 
in which the parameter $\nu^2$ plays a role of the central charge. 
This is not surprising since such a nonlinearity characterizes
the symmetry algebras of systems with a conserved 
Laplace-Runge-Lenz vector.
The nonlinearity originates from the particular form
of the integral (\ref{Xi0})\,: it is cubic in the phase space 
variables\footnote{A nonlinear  superalgebraic
structure associated with the squared total angular momentum
appears also in the supersymmetrized charge-monopole system, see 
\cite{PlMon,DeJMPH,PnSU,LePl}.}
$\Pi_i$, $r_j$ and $\xi_a$.

 The Poisson brackets between $\Xi_i$ and $\Xi_j$ 
can also be computed by using  
relations (\ref{SXi}) and employing the graded
 Jacobi identities.
Again, we get a nonlinear 
(quadratic in the integrals $\mathcal{J}_i$ and 
$\mathcal{S}^+_j$)
Poisson bracket relation,
\begin{equation}\label{XiiXij}
    \{\Xi_i,\Xi_j\}=i\delta_{ij}\left(
    \nu^2-(\vec{\mathcal{J}}-2\vec{\mathcal{S}}^+)^2\right)
    +4i\mathcal{S}^+_i\mathcal{S}^+_j
    -2i(\mathcal{S}^+_i\mathcal{J}_j+
    \mathcal{S}^+_j\mathcal{J}_i)\,. 
\end{equation}
The scalar integral $\Xi_0$ Poisson-commutes with 
two other scalar Grassmann-odd integrals,
\begin{equation}
    \{\Theta_0,\Xi_0\}=\{\Omega_0,\Xi_0\}=0\,.
\end{equation}
On the other hand, we have nontrivial Poisson bracket relations
\begin{equation}
    \{\Theta_i,\Xi_0\}=-\{\Theta_0,\Xi_i\}=
    i\mathcal{G}_i\,,\qquad
    \{\Omega_i,\Xi_0\}=
    -\{\Omega_0,\Xi_i\}=
    i\mathcal{R}_i\,,
\end{equation}
\begin{equation}
    \{\Xi_i,\Theta_j\}=-i\epsilon_{ijk}\mathcal{G}_k\,,\qquad
    \{\Xi_i,\Omega_j\}=-i\epsilon_{ijk}\mathcal{R}_k\,.
\end{equation}    
This provides us with  a  generalization of the 
Laplace-Runge-Lenz vector integral (\ref{Gvec0})
 and its associated dynamical integral (\ref{Rint}),
\begin{equation}\label{LRLdef+}
    \vec{\mathcal{G}}= \vec{\Pi}\times
    (\vec{\mathcal{J}}-\vec{\mathcal{S}}^+)
    +\vec{\Pi}\times
    \vec{\mathcal{S}}^- +\frac{2\nu}{r}
    \left(\vec{\mathcal{S}}^--\frac{\vec{r}}{r}(\vec{r}\cdot
    \vec{\mathcal{S}}^-)\right),
\end{equation}
\begin{equation}\label{Ndef}
    \vec{\mathcal{R}}=\vec{r}\times (\vec{\mathcal{J}}
    -\vec{\mathcal{S}}^+ +\vec{\mathcal{S}}^-)-
    2\vec{\mathcal{G}}t\,.
\end{equation}
The supersymmetrized Laplace-Runge-Lenz vector 
(\ref{LRLdef+}) can be written in the form
$$  
 \vec{\mathcal{G}}= \vec{\Pi}\times
    \vec{\mathcal{Y}}^--\frac{i}{2}\vec{\xi}\times
    \vec{\Theta}+i\frac{\nu}{2r^3}\vec{\rho}\times
    \vec{\rho}\,.
$$
It Poisson-commutes with the chiral spin vector and supercharges $\Theta_a$,
\begin{equation}\label{APoiss}
    \{\mathcal{G}_i,\mathcal{S}^+_j\}=
    \{\mathcal{G}_i,\Theta_a\}=0\,,
\end{equation}
while the brackets with the conformal symmetry generators 
are
\begin{equation}\label{APoiss}
    \{\mathcal{H},\mathcal{G}_i\}=0\,,\qquad
    \{\mathcal{D},\mathcal{G}_i,\}=\mathcal{G}_i\,,\qquad
    \{\mathcal{K},\mathcal{G}_i,\}=2\mathcal{R}_i\,.
\end{equation}
For the dynamical vector integral $\mathcal{R}_i$ we 
have
\begin{equation}\label{ROmegaS}
  \{\mathcal{R}_i,\mathcal{S}^+_j\}=
    \{\mathcal{R}_i,\Omega_a\}=0\,,
\end{equation}
and  in addition
\begin{equation}\label{NDHKS}
\{\mathcal{H},\mathcal{R}_i\}=-2\mathcal{G}_i\,,\qquad
    \{\mathcal{D},\mathcal{R}_i,\}=-\mathcal{R}_i\,,\qquad
    \{\mathcal{K},\mathcal{R}_i,\}=0\,.
\end{equation}
We also have Lie type Poisson bracket relations
\begin{equation}\label{OmegaG}
    \{\Omega_0,\mathcal{G}_i\,\}=\Xi_i\,,\qquad
    \{\Omega_i,\mathcal{G}_j,\}=
    \epsilon_{ijk}\Xi_k-\delta_{ij}\Xi_0\,,
\end{equation}
\begin{equation}
    \{\Theta_0,\mathcal{R}_i\,\}=-\Xi_i\,,\qquad
    \{\Theta_i,\mathcal{R}_j,\}=
    -\epsilon_{ijk}\Xi_k+\delta_{ij}\Xi_0\,.
\end{equation}
It is worth noting here, that although in 
(\ref{OmegaG}) the $\Xi_a$ are generated
via Poisson brackets of the \emph{dynamical} 
odd integrals $\Omega_a$ 
and  the true 
 integrals $\mathcal{G}_i$, 
they are  true Grassmann integrals.
This happens since the  time-dependent term 
 $-2t\Theta_a$ in $\Omega_a$  Poisson-commutes 
with $\mathcal{G}_i$.

The Poisson brackets of $\mathcal{G}_i$ and 
$\mathcal{R}_i$ with $\Xi_a$
are quadratic polynomials in the integrals,
\begin{equation}
    \{\Xi_0,{\mathcal{G}_i}\}=
    (\vec{\mathcal{J}}\times\vec{\Theta})_i\,,\qquad
     \{\Xi_0,{\mathcal{R}}_i\}=
   (\vec{\mathcal{J}}\times  \vec{\Omega})_i\,,
\end{equation}
\begin{equation}
    \{\mathcal{G}_i,\Xi_j\}=\delta_{ij}
    (\vec{\mathcal{J}}-2\vec{\mathcal{S}}^+)\cdot\vec{\Theta}-
    \Theta_0\epsilon_{ijk}\mathcal{J}_k+
    2\mathcal{S}^+_i\Theta_j-
    \Theta_i\mathcal{J}_j
    \,,
\end{equation}
\begin{equation}
    \{\mathcal{R}_i,\Xi_j\}=\delta_{ij}
    (\vec{\mathcal{J}}-2\vec{\mathcal{S}}^+)\cdot\vec{\Omega}-
    \Omega_0\epsilon_{ijk}\mathcal{J}_k+
    2\mathcal{S}^+_i\Omega_j-
    \Omega_i\mathcal{J}_j
    \,.
\end{equation}
The Poisson brackets between the integrals
$\mathcal{G}_i$ and $\mathcal{R}_i$
are also quadratic,
\begin{equation}\label{GiGj}
    \{\mathcal{G}_i,\mathcal{G}_j\}=
    -\mathcal{H}\epsilon_{ijk}
    \mathcal{J}_k
    -\frac{i}{2}\epsilon_{ijk}\left(
    \vec{\Theta}\times\vec{\Theta}\right)_k\,,
\end{equation}
\begin{equation}\label{RiRj}
    \{\mathcal{R}_i,\mathcal{R}_j\}=
    -\mathcal{K}\epsilon_{ijk}
    \mathcal{J}_k
    -\frac{i}{2}\epsilon_{ijk}\left(
    \vec{\Omega}\times\vec{\Omega}\right)_k\,,
\end{equation}
\begin{eqnarray}
    \{\mathcal{G}_i,\mathcal{R}_j\}&=&\delta_{ij}
    \left(\nu^2-2\vec{\mathcal{J}}^2
    +4
    \vec{\mathcal{J}}\cdot\vec{\mathcal{S}}^+
    +i
    \vec{\Omega}\cdot\vec{\Theta}\right) +\frac{i}{2}(\Theta_i\Omega_j
    +\Theta_j\Omega_i)+\mathcal{J}_i\mathcal{J}_j
    -2(\mathcal{S}_i^+\mathcal{J}_j+
    \mathcal{S}_j^+\mathcal{J}_i)\nonumber\\
    &-&
    \epsilon_{ijk}\left( \mathcal{D}\mathcal{J}_k+\frac{i}{2}
    (\vec{\Theta}\times\vec{\Omega})_k\right)
    \,.\label{AiNj}
\end{eqnarray}
With these relations we obtain a closed, nonlinear
(quadratic)
superconformal algebra extended by
the Laplace-Runge-Lenz vector $\vec{\mathcal{G}}$,
the associated Grassmann-even dynamical vector integral $\vec{\mathcal{R}}$,
and  by the Grassmann-odd integrals $\Xi_a$.

Let us now discuss shortly some aspects of the 
quantum version of the  described supersymmetric structure,
in which the Poisson brackets between the Grassmann-odd generators
become anticommutators of the corresponding quantum
fermionic operators, while the  brackets between the Grassmann-even 
with Grassmann-even or Grassmann-odd generators  become
commutators between  their quantum counterparts.

The quantum analog of the Hamiltonian (\ref{Hsefldyon+}) is
the matrix $4\times 4$ operator having a block-diagonal form,
\begin{equation}\label{hHc}
    \hat{\mathcal{H}}=
    \hat{\vec{\Pi}}^2+\frac{\nu^2}{r^2}
    -\frac{2\nu}{r^3}\mathcal{T}_- (1\otimes \vec{\sigma}\cdot\vec{r}\,)
    =
\left(
\begin{array}{cc}
    \hat{H}_+ & 0   \\
    0 &  \hat{H}_-
\end{array}
\right)
\,,
\end{equation}
where the $2\times 2$ Hamiltonians are
$\hat{H}_+=\hat{\vec{\Pi}}^2+\nu^2/r^2$
and
$\hat{H}_-=\hat{\vec{\Pi}}^2+\nu^2/r^2
-2\nu(\vec{\sigma}\cdot\vec{r}\,)/r^3.$
Note that the chiral operator $\hat{H}_+$ 
is proportional to the $2\times 2$ identity matrix.
This can be attributed to the self-duality of the 
dyon field. The  conserved total angular momentum operator
$\hat{\vec{\mathcal{J}}}=\hat{\vec{J}}+
\hat{\vec{\mathcal{S}}}$,
contains the spinless part 
$\hat{\vec{J}}=\vec{r}\times\hat{\vec{\Pi}}-
\nu\vec{n}$ and  
the spin operator 
$\hat{\vec{\mathcal{S}}}=1\otimes\frac{1}{2}\vec{\sigma}$.
By standard arguments, the parameter $\nu$  
undergoes the Dirac quantization: at the quantum level
it can take only integer or 
half-integer values, i.e. $\nu=n/2$, $n\in\Z$ 
\cite{GoddOl,Dirac}.

The operator $\hat{H}_+$ represents two identical copies
of the  Hamiltonian operator of the
spinless system discussed in the previous section, and
the integral nature of the chiral spin operator
$\hat{\vec{\mathcal{S}}}^+=\mathcal{T}_+
(1\otimes\frac{1}{2}\vec{\sigma})$
is then obvious. The diagonal operator 
$\hat{{\mathcal{S}}}^+_3$  distinguishes  
the upper and lower components 
of this doubled spinless system, while the operators
$\hat{{\mathcal{S}}}^+_1\pm i\hat{{\mathcal{S}}}^+_2$
transform them one into another
in an obvious way.
The  operator $\hat{H}_-$
can be interpreted as the Pauli type Hamiltonian of the 
charged spin-$1/2$ particle 
of gyromagnetic ratio  $4$ in the combined field
of the magnetic monopole and scalar potential $\nu^2/r^2$.

The scalar and vector quantum supercharges for
the extended system (\ref{hHc}) have
a block-antidiagonal form
\begin{equation}\label{hatTh}
	\hat{\Theta}_a=\frac{1}{\sqrt{2}}
\left(
\begin{array}{cc}
0  &  \hat{Q}_a    \\
\hat{Q}^\dagger_a  & 0   
\end{array}
\right),
\end{equation}
where
\begin{equation}
	\hat{Q}_0=-i\vec{\sigma}\cdot
	\hat{\vec{\Pi}}+\frac{\nu}{r}\,,\qquad
	\hat{\vec{Q}}=-i\vec{\sigma}
	\hat{Q}_0=-i\big(\hat{\vec{\Pi}}\times
	\vec{\sigma}+\frac{\nu}{r}\vec{\sigma}\big)-
	\hat{\vec{\Pi}}\,.
\end{equation}
They commute with the Hamiltonian (\ref{hHc}) and anticommute
with the diagonal 
grading operator $\Gamma$ in (\ref{Gamma}),
which itself is an additional quantum integral of motion 
of the bosonic nature.
Note that $\hat{Q}_a$ and 
$\hat{Q}^\dagger_a$ are the Darboux 
intertwining operators\,:
\begin{equation}\label{Darboux}
	\hat{Q}_a \hat{H}_-=
	\hat{H}_+\hat{Q}_a\,,\qquad 
	\hat{Q}_a^\dagger \hat{H}_+=
	\hat{H}_-\hat{Q}_a^\dagger\,.
\end{equation}
These relations are equivalent to the condition of commutativity 
of supercharges $\hat{\Theta}_a$ and 
$i\Gamma\hat{\Theta}_a$ with $\hat{\mathcal{H}}$.
As in the case of the one-dimensional  supersymmetric quantum mechanics,
they allow us to relate the eigenstates  $\psi_+$ and $\psi_-$
of the Hamiltonians $\hat{H}_+$ and  $\hat{H}_-$ 
of the quantum subsystems,
$\psi_+\propto \hat{Q}_a\psi_-$,  
$\psi_-\propto \hat{Q}^\dagger_a\psi_+$.

The superconformal generators $\hat{\Omega}_a$ have a 
block-antidiagonal form similar to that in (\ref{hatTh}), 
with   $\hat{Q}_0$ replaced by
$-i\vec{\sigma}\cdot\vec{r}-2
\hat{Q}_0t$, and $\hat{\vec{Q}}$  replaced by 
$-i\vec{r}\times\vec{\sigma}-\vec{r}-2\hat{\vec{Q}}t$.
The quantum analogs $\hat{\Xi}_a$ of the supercharges 
(\ref{Xi0}) and (\ref{Xii}) have the same 
 block-antidiagonal form, 
but with the scalar operator $\hat{Q}_0$   
replaced by 
$i(\vec{\sigma}\cdot\hat{\vec{L}}+1)$,
and the vector operator $\hat{\vec{Q}}$ replaced by
$
-(\hat{\vec{L}}+\vec{\sigma})-i\hat{\vec{L}}\times
\vec{\sigma}$, where 
 $\hat{\vec{L}}=\vec{r}\times\hat{\vec{\Pi}}$.
Note that in the anticommutator 
$[\hat{\Xi}_0,\hat{\Xi}_0]_+=\hat{\vec{\mathcal{J}}}^2
-\nu^2+1/4$, there appears  a 
quantum correction term $\hbar^2/4$. 

The Laplace-Runge-Lenz vector operator 
of the supersymmetric quantum system is
\begin{equation}
    \hat{\vec{\mathcal{G}}}=
    \hat{\vec{G}}
+
\mathcal{T}_-\cdot 1\otimes \left(\hat{\vec{\Pi}}\times
\vec{\sigma}+\frac{\nu}{r}
\left(\vec{\sigma}-
\vec{n}(\vec{\sigma}\cdot\vec{n})
\right)\right)\,,
\end{equation}
where  $\hat{\vec{G}}$ is a Hermitian spinless 
Laplace-Runge-Lenz vector,
$$
\hat{\vec{G}}=\frac{1}{2}\big(\hat{\vec{\Pi}}\times
\hat{\vec{J}}-\hat{\vec{J}}\times
\hat{\vec{\Pi}}\big)=-\hat{\vec{J}}\times
\hat{\vec{\Pi}}+i\hat{\vec{\Pi}}\,.
$$
It commutes with the supercharges $\hat{\Theta}_a$, 
and thus with the Hamiltonian $\hat{\mathcal{H}}$.
The quantum analog of the related, dynamical vector integral (\ref{Ndef})
is
\begin{equation}
	\hat{\vec{\mathcal{R}}}=\frac{1}{2}\big({\vec{r}}\times
\hat{\vec{J}}-\hat{\vec{J}}\times\vec{r}\,\big) -\mathcal{T}_-
\big(1\otimes \vec{\sigma}\times\vec{r}\,\big) -
2\hat{\vec{G}}t\,.
\end{equation}
In correspondence with the classical properties,
it
satisfies the commutation relation
$[\hat{\vec{\mathcal{R}}},\hat{\mathcal{H}}]=2i\hat{\vec{\mathcal{G}}}$.

\section{Summary, concluding remarks and outlook}

Let us summarize the obtained nonlinear superalgebraic 
structure of the spinning 
charged  particle in the background of the self-dual dyon. \vskip0.1cm

A very particular role in the supersymmetric  structure 
is played by the chiral spin vector $\vec{\mathcal{S}}^+$, whose 
origin  is rooted in the self-dual nature of the background electromagnetic 
field. 
This bosonic, Grassmann-even vector integral commutes with all other 
basic bosonic integrals, except the total angular momentum vector
$\vec{\mathcal{J}}$. 
On the other hand, all the fermionic, Grassmann-odd integrals 
are grouped into the three irreducible representations with respect 
to the Poisson bracket action, or, commutator in the quantum case,
of  $\vec{\mathcal{S}}^+$ on them. 
The rotational symmetry associated with 
$\vec{\mathcal{J}}$ is due to a spherical
symmetry of the magnetic  
and dual to it electric fields
of the dyon.
The sets of the  integrals 
$\mathcal{Y}_i^-=\mathcal{J}_i-\mathcal{S}^+_i$ 
and $\mathcal{Y}_i^+=\mathcal{S}^+_i$
generate the $so(3)\oplus so(3)=so(4)$ Lie subalgebra.

One can distinguish the three  entangled supersymmetry
substructures in the system,
each of which can be related to its corresponding 
basic bosonic integral.  

\vskip0.1cm

{\bf i)}   The  
Hamiltonian $\mathcal{H}$ is a supersymmetric 
 generalization
of the spinless  Hamiltonian (\ref{Hspec}), which 
possesses conformal symmetry and  reveals
a hidden partially free dynamics.  
The peculiar dynamics of the spinless system
is encoded  in the presence
of  the conserved Laplace-Runge-Lenz vector  $\vec{G}$ 
and the associated
dynamical integral $\vec{R}$, which 
can be related to the deformed  Galilei symmetry. 
Quantum mechanically, the square root of 
$\hat{\mathcal{H}}$
is the scalar
supercharge $\hat{\Theta}_0$, whose classical 
Grassmann-odd analog $\Theta_0$ Poisson-commutes
with itself for $-i\mathcal{H}$.
The Poisson bracket of $\Theta_0$  with the chiral spin vector 
integral
$\vec{\mathcal{S}}^+$ generates three more integrals, $\Theta_i$,
which form a vector $\vec{\Theta}$ with respect to the total angular momentum
integral $\vec{\mathcal{J}}$. Each $\Theta_i$, like $\Theta_0$, is a
square root of   $\mathcal{H}$: together these four supercharges $\Theta_a$
generate the $N=4$ supersymmetry:
$\{\Theta_a,\Theta_b\}=-i\delta_{ab}\mathcal{H}$,
$\{\Theta_a,\mathcal{H}\}=0$.
The Poisson bracket of $\Theta_i$ with $\mathcal{S}^+_j$
is a linear combination of $\Theta_0$ and $\Theta_k$.
On the quantum level the interplay between self-duality
and extended supersymmetry has been emphasized in a more
general context (without further assumptions) in 
\cite{KirLanWip},
see also \cite{CromRit} and \cite{IvaLech}.

\vskip0.2cm

{\bf ii)} The dynamical integral $\mathcal{K}$, which
explicitly depends on time,
generates the special conformal transformations.
Its bracket with the Hamiltonian $\mathcal{H}$ produces 
the  generator of dilatations $\mathcal{D}$ as a further 
dynamical integral. The integrals $\mathcal{H}$, 
$\mathcal{K}, \mathcal{D}$ 
Poisson-commute with the integrals $\mathcal{J}_i$ and
$\mathcal{S}^+_i$, and  generate a $so(2,1)$ symmetry. 
The classical analog of the square root of
the quantum operator $\hat{\mathcal{K}}$ 
corresponds to the Grassmann-odd
dynamical scalar integral $\Omega_0$. The Poisson brackets of
$\Omega_0$ with the integrals $\mathcal{S}^+_i$ generate three more
dynamical integrals $\Omega_i$, which form a vector with respect
to $\vec{\mathcal{J}}$. The set $\Omega_a$, $a=0,i$, has the same
transformational properties with respect to $\mathcal{S}^+_i$
as the Grassmann odd integrals $\Theta_a$. 
The Grassmann-odd dynamical integrals $\Omega_a$ 
Poisson commute with $\mathcal{K}$, and together they generate 
the sub-superalgebra, similar to that 
generated by $\Theta_a$ and $\mathcal{H}$:
$\{\Omega_a,\Omega_b\}=-i\delta_{ab}\mathcal{K}$,
$\{\Omega_a,\mathcal{K}\}=0$.

The dynamical integrals $\Omega_a$ are eigenstates 
of $\mathcal{D}$ with eigenvalue $-1$, which means $\{\mathcal{D},
\Omega_a\}=-\Omega_a$.
Similarly $\mathcal{K}$ is an eigenstate with
eigenvalue $-2$ and $\mathcal{H}$ is an eigenstate
with eigenvalue $+2$. Accordingly one finds that the
conserved supercharges  $\Theta_a$ are eigenstates
of $\mathcal{D}$ of eigenvalue $+1$.

The superalgebraic structures outlined in i) and ii)
are entangled via via Poisson brackets. The bracket of $\mathcal{H}$ 
with the dynamical integrals
 $\Omega_a$ produces the integrals $\Theta_a$,
 while the  bracket of $\mathcal{K}$
 with the integrals $\Theta_a$ generates $\Omega_a$.
 The Poisson brackets between the supercharges $\Theta_a$ 
 and superconformal charges
 $\Omega_a$ produce linear combinations 
 of $\mathcal{D}$, $\mathcal{J}_i$
and $\mathcal{S}^+_i$, from which these
even integrals
can be completely reconstructed.

The usual integrals $\mathcal{H}$, $\mathcal{Y}_i^\pm$, $\Theta_a$ 
(which do not depend explicitly on time)
together with the dynamical integrals $\mathcal{K}$, $\mathcal{D}$,
and $\Omega_a$ generate the exceptional simple
Lie superconformal algebra $D(2,1;\alpha)$ with $\alpha=1/2$. 
This Lie superalgebra of order $17$ has the quadratic Casimir 
$\mathcal{C}=\frac{1}{4}(\mathcal{H}\mathcal{K}-
\mathcal{D}^2)+\frac{1}{2}
\vec{\mathcal{Y}}^-{}^2
-\frac{3}{2}\vec{\mathcal{Y}}^+{}^2
+\frac{i}{2}
\Theta_a\Omega_a$.
To the last Grassmann-even nilpotent term corresponds 
the quantum operator $\frac{i}{4}[\hat{\Theta}_a,\hat{\Omega}_a]
= \hat{\mathcal{L}}_\sigma+3/2$ with $\hat{\mathcal{L}}_\sigma\equiv
\mathcal{T}_-[1\otimes(\hat{\vec{L}}+\nu\vec{n})\cdot
\vec{\sigma}]$ which is a nontrivial integral
for the spin-1/2 subsystem $\hat{H}_-$. The quantum bosonic integral
$\hat{\mathcal{L}}_\sigma$  satisfies
the quadratic  relation
$\hat{\mathcal{L}}_\sigma(
\hat{\mathcal{L}}_\sigma+2)
=\mathcal{T}_-\big(\hat{\vec{\mathcal{J}}}^2-3/4\big)$.

\vskip0.2cm
{\bf iii)}  Similarly as for $\hat{\mathcal{H}}$ and 
$\hat{\mathcal{K}}$, there exists a anti-diagonal
fermionic square root for the conserved
operator $\hat{\vec{\mathcal{J}}}^2$. Its classical analog
is the Grassmann-odd scalar
integral $\Xi_0$, which
satisfies the Poisson bracket relation
$\{\Xi_0,\Xi_0\}=-i(\vec{\mathcal{J}}^2-\nu^2)$.
The bracket of $\Xi_0$ with the chiral spin vector $\vec{\mathcal{S}}^+$
produces the Grassmann odd vector integral $\vec{\Xi}$,
and the set $\Xi_a$ is transformed by $\mathcal{S}^+_i$
in the same way as the integrals $\Theta_a$  and $\Omega_a$.
However, $\Xi_i$ has nonzero Poisson brackets
with $\Xi_0$, and the brackets of 
$\Xi_a$ with $\Xi_b$ turn out to be quadratic
in the total angular momentum $\mathcal{J}_i$ 
and the chiral spin vector $\mathcal{S}^+_i$. 
On the other hand,
in contrast with  the Grassmann-odd integrals $\Theta_a$ and $\Omega_a$,
the integrals $\Xi_a$,  like $\mathcal{J}_i$ 
and  $\mathcal{S}^+_i$,  Poisson-commute with the $so(2,1)$
generators $\mathcal{K}$ and $\mathcal{D}$,
which means that they are $so(2,1)$ scalars.

The Poisson brackets of $\Xi_a$ with $\Theta_b$ generate
 the Laplace-Runge-Lenz  vector integral $\mathcal{G}_i$. 
 Analogously, the brackets of
$\Xi_a$ with $\Omega_b$ produce the dynamical vector integral
$\mathcal{R}_i$ associated with $\mathcal{G}_i$, 
$\frac{\partial}{\partial t}\mathcal{R}_i=\mathcal{G}_i$.
 In correspondence with
this, $\mathcal{G}_i$ and $\mathcal{R}_i$ are eigenvectors of
$\mathcal{D}$ with eigenvalues $+1$ and $-1$, respectively. 
The Poisson bracket of $\mathcal{G}_i$ with $\mathcal{K}$ leads to
$\mathcal{R}_i$, while $\mathcal{R}_i$ Poisson-commutes with
$\mathcal{K}$. The Hamiltonian $\mathcal{H}$ acts on these vector
integrals in  the opposite way: it Poisson-commutes with
$\mathcal{G}_i$, and its bracket with the dynamical integral
$\mathcal{R}_i$ yields the  integral $\mathcal{G}_i$. 
These bosonic vector integrals 
Poisson-commute with the chiral spin vector
$\mathcal{S}^+_i$. At the same time, 
the vector $\mathcal{G}_i$ Poisson-commutes
with $\Theta_a$, while its brackets with $\Omega_a$ produce
integrals $\Xi_a$. Analogously, the dynamical integral
$\mathcal{R}_i$ Poisson commutes with $\Omega_a$, and generates
the Grassmann-odd integrals $\Xi_a$ via the brackets with $\Theta_a$. The
Poisson brackets of $\mathcal{G}_i$ and $\mathcal{R}_i$ with
$\Xi_a$,  and between $\mathcal{G}_i$ and $\mathcal{R}_i$
yields quadratic expressions of the other integrals. So, the extension 
of the set of the generators of the superconformal 
symmetry $D(2,1;\alpha=1/2)$ by the 
Grassmann-odd and Grassmann-even integrals  
$\Xi_a$ and $\mathcal{G}_i,\mathcal{R}_i$ 
transforms the Lie superalgebra into a 
nonlinear, quadratic superalgebra.

It is worth noting that starting  with any one of the set of 
the ten integrals
$\Xi_a$, $\mathcal{G}_i$  and $\mathcal{R}_i$, one can  generate
the complete set of integrals by taking Poisson brackets with the
generators of $D(2,1;\alpha=1/2)$,   for instance,
$$\Xi_0 \xmapsto{\mathcal{\,\,S}^+_i\,\,}\Xi_i
\xmapsto{\,\,\Theta_j\,\,}
\mathcal{G}_i\xmapsto{\,\,\,\mathcal{K}\,\,\,}\mathcal{R}_i
\xmapsto{\,\,\Theta_j\,\,}\Xi_0,\Xi_i\,.$$
Then, as the minimal set of the integrals generating 
all the quadratic superalgebra via the 
Poisson brackets one can take
one of the Grassmann-odd integrals $\Theta_a$, 
one of the Grassmann-odd dynamical integrals 
$\Omega_b$, two components of the chiral spin vector 
$\vec{\mathcal{S}}^+$,
and finally one of the odd or even integrals 
$\Xi_a$ or $\mathcal{J}_i, \mathcal{R}_i$.
\vskip0.1cm

Quantum mechanically the system has the additional integral
$\Gamma=\gamma_5=-\gamma_0\gamma_1\gamma_2
\gamma_3=\tau_3\otimes 1$. 
It commutes with quantum analogs of all the Grassmann-even integrals,
but anticommutes  with  the quantum analogs of all the 
Grassmann-odd integrals.
It is identified as the $\Z_2$-grading operator of the
quantum supersymmetric structure. The multiplication of
any of the fermionic Hermitian generators $\hat{\Phi}_\mu$ 
from the set ($\hat{\Theta}_a$,  $\hat{\Omega}_a$, $\hat{\Xi}_a$)
by $i\gamma_5$ gives a new Hermitian fermionic integral, and 
in an algebraic way doubles the  set of them.  As a result, 
instead of the nonlinear, quadratic 
superalgebra consisting of 12 fermionic integrals, we end up 
with a superalgebra with 24 fermionic generators satisfying
the commutation relations 
$[\Gamma,\hat{\Phi}_{s\mu}]=-2i\epsilon_{ss'}
\hat{\Phi}_{s'\mu}$, where 
$s,s'=1,2$, $\hat{\Phi}_{1\mu}=\hat{\Phi}_{\mu}$
and $\hat{\Phi}_{2\mu}=i\Gamma\hat{\Phi}_{1\mu}$.
The linear combinations 
$\hat{\Phi}_{\pm \mu}=\frac{1}{2}(\hat{\Phi}_{1\mu}\mp i
\hat{\Phi}_{2\mu})=
\mathcal{T}_\pm\hat{\Phi}_\mu$ with
$(\hat{\Phi}_{+\mu})^\dagger=
\hat{\Phi}_{-\mu}$, define chiral fermionic generators, see 
Eq. (\ref{hatTh}), and the integrals $\hat{\Xi}_{\pm a}$ are Darboux
generators that intertwine the spin-independent (doubled)
Schr\"odinger operator $\hat{H}_+$ and spin-dependent 
Pauli operator $\hat{H}_-$.
This property has a natural explanation if one views
$\hat{H}_\pm$ as parts of a squared Euclidean Dirac operator 
for selfdual background fields \cite{KirLanWip}.
Analogously, the  $\hat{\Omega}_{\pm a}$ are
Darboux intertwiners for the operators $\hat{\mathcal{K}}_+$ and 
$\hat{\mathcal{K}}_-$,
$\hat{\mathcal{K}}_\pm=\mathcal{T}_\pm
\hat{\mathcal{K}}$,
which are generators of the special conformal symmetries
of the corresponding spinless and spin-1/2 quantum
subsystems.

On the other hand, at the classical level the quantity
$-2\xi_0\xi_1\xi_2\xi_3=\frac{1}{3}\vec{\mathcal{S}}^+{}^2$
corresponds to the operator $\gamma_5$. This, however,
is an Grassmann-even integral,  whose quantum analog
includes a multiplicative factor
$\hbar^2$ in contrast with the quantum analog 
of $\xi_a$ to be proportional to $\sqrt{\hbar}$.
A multiplication of this nilpotent Grassmann-even integral 
by any of the classical Grassmann-odd 
 integral  produces zero,
 and we have no classical analog for the quantum fermionic 
 integrals $\hat{\Phi}_{2\mu}$.
  Alternatively, as a classical analog of $\gamma_5$,
 one could  try to introduce  an independent Grassmann scalar variable
 $\xi_5$ with the only nontrivial Poisson bracket $\{\xi_5,\xi_5\}=-i$.
In this case multiplication of classical integrals
 $\Theta_a$, $\Omega_a$ and $\Xi_a$ by $i\xi_5$ will give new
 classically nontrivial integrals. However, they will be of
 the Grassmann-even, bosonic  nature. The described problems
 associated with the introduction of the 
 classical analog for the matrix $\gamma_5$ 
 are well known  and were discussed in a different context in
 the  literature \cite{gamma5}.

\vskip0.1cm

Based on relations (\ref{VGH}) and (\ref{JSJ})
for the spinless bosonic case,
one could assume that 
by a suitable nonlinear (non-local at the quantum level)
redefinition of the  
even, ${\mathcal{G}}_i$  and ${\mathcal{R}}_i$,
and the odd, $\Xi_a$,  generators of the
nonlinear superconformal symmetry
it can be reduced  to some linear
superextension of the $D(2,1;\alpha=1/2)$.
Let us show that such a linearization, however, 
is impossible.  

Nonlinear (quadratic) algebraic relations (\ref{GiGj})
and (\ref{RiRj}) are linearized by a redefinition
$\mathcal{G}_i\rightarrow \breve{\mathcal{G}}_i$,
$\mathcal{R}_i\rightarrow \breve{\mathcal{R}}_i$,
\begin{equation}\label{breveGR}
	\breve{\mathcal{G}}_i=\frac{1}{\sqrt{\mathcal{H}}}\,\mathcal{G}_i
	+\frac{1}{2\mathcal{H}}(\vec{\Theta}\times\vec{\Theta})_i\,,\qquad
	\breve{\mathcal{R}}_i=\frac{1}{\sqrt{\mathcal{K}}}\,\mathcal{R}_i
	+\frac{1}{2\mathcal{K}}(\vec{\Omega}\times\vec{\Omega})_i\,.
\end{equation}
These vectors satisfy, particularly,  the Lie-Poisson algebraic relations
\begin{equation}
	\{\breve{\mathcal{G}}_i,\breve{\mathcal{G}}_j\}=
	-\epsilon_{ijk}\mathcal{J}_k\,,\qquad
	\{\breve{\mathcal{G}}_i,\Theta_j\}=i\epsilon_{ijk}\Theta_k\,,
	\qquad
	\{\breve{\mathcal{G}}_i,\mathcal{H}\}=
	\{\breve{\mathcal{G}}_i,\mathcal{D}\}=
	\{\breve{\mathcal{G}}_i,\Theta_0\}=0\,,
\end{equation}
and
\begin{equation}
	\{\breve{\mathcal{R}}_i,\breve{\mathcal{R}}_j\}=
	-\epsilon_{ijk}\mathcal{J}_k\,,\qquad
	\{\breve{\mathcal{R}}_i,\Omega_j\}=i\epsilon_{ijk}\Omega_k\,,
	\qquad
	\{\breve{\mathcal{R}}_i,\mathcal{K}\}=
	\{\breve{\mathcal{R}}_i,\mathcal{D}\}=
	\{\breve{\mathcal{R}}_i,\Omega_0\}=0\,.
\end{equation}
 One can also redefine $\Xi_a\rightarrow\breve{\Xi}_a$,
\begin{equation}
	\breve{\Xi}_0=\frac{\Xi_0}{\sqrt{\vec{\mathcal{J}}^2-\nu^2}}\,,\qquad
	\breve{\Xi}_i=\frac{\Xi_i}{\sqrt{\vec{\mathcal{J}}^2-\nu^2}}
	-\frac{2\breve{\Xi}_0}{\vec{\mathcal{J}}^2-\nu^2}(\vec{\mathcal{J}}\times
	\vec{\mathcal{S}}^+)_i\,.
\end{equation}
These integrals satisfy with the $\mathcal{S}^+_i$
the same relations as the initial odd integrals $\Xi_a$,
 $\{\mathcal{S}^+_i,\breve{\Xi}_0\}=\frac{1}{2}\breve{\Xi}_i$,
$\{\mathcal{S}^+_i,\breve{\Xi}_j\}=\frac{1}{2}(\epsilon_{ijk}\breve{\Xi}_k
-\delta_{ij}\breve{\Xi}_0)$, but instead of the nonlinear
Poisson bracket relations (\ref{Xi0Xi0}), 
(\ref{Xi0Xii}) and (\ref{XiiXij}) we get
\begin{equation}
	\{\breve{\Xi}_a,\breve{\Xi}_b\}=-i\delta_{ab}\,.
\end{equation}
These last Poisson brackets
 can be treated as a superalgebraic relation
with central charge $1$.
The brackets
\begin{equation}
	\{\Theta_i,\breve{\Xi}_0\}=-\{\Theta_0,\breve{\Xi}_i\}=
	i\frac{\mathcal{G}_i}{\sqrt{\vec{\mathcal{J}}^2-\nu^2}}
	-\frac{(\vec{\mathcal{J}}\times
	\vec{\Theta})_i}{\vec{\mathcal{J}}^2-\nu^2}
	\breve{\Xi}_0
\end{equation}
together with Eq. (\ref{breveGR})
show nevertheless  that the redefined integrals generate a more complicated,
non-polynomial superalgebra (which 
is of a non-local nature at the quantum level). 
Analogous non-polynomial structures appear 
in the Poisson brackets of 
$\Theta_a$ and $\mathcal{H}$ with $\breve{\mathcal{R}}_i$,
and in those of $\Omega_a$ and $\mathcal{K}$ 
with $\breve{\mathcal{G}}_i$. The brackets of 
$\breve{\mathcal{G}}_i$ with $\breve{\mathcal{R}}_j$
are also of a non-polynomial form in other integrals.
This shows the impossibility of the linearization 
of the obtained quadratic extension of the superconformal
symmetry $D(2,1;\alpha=1/2)$.

\vskip0.1cm

The model we considered possesses only  the continuous
spectrum  with $E>0$. It is known that in the quantum  
Kepler problem the hidden symmetry associated with the 
dynamical Laplace-Runge-Lenz vector integral
not only determines the energy levels completely,
but also the phase shifts   \cite{Zwanz2}. 
One could naturally expect that the hidden supersymmetry 
we discussed  here may also be helpful for
analyzing the scattering characteristics of the 
quantum system. Investigation of this problem 
is outside of the scope of the 
present work and deserves a separate consideration. 

\vskip0.1cm

The dynamics of the 
boson (spinless) sector of the system we studied  
corresponds to  the special bosonic
dynamics of the $D(2,1;\alpha)$ model with 
$\alpha^2=1/4$  investigated by 
Ivanov, Krivonos and Lechtenfeld in \cite{Ivanov+}. 
The approach  of \cite{Ivanov+} is
completely different, however, that 
allowed the authors to obtain the 
general $D(2,1;\alpha)$ models. 
Namely, in  \cite{Ivanov+}  the $N=4$ 
superconformal mechanics
realizing the  $D(2,1;\alpha)$ symmetry
with \emph{arbitrary} values of the parameter $\alpha$ 
was constructed 
in the $N=4$, $d=1$ superspace. 
It was shown then that  
the bosonic sector of the system 
describes a conformally invariant nonlinear sigma model,
which at the two particular 
values $\alpha=+1/2$ and $\alpha=-1/2$ 
reduces to the system (\ref{Hspec})\footnote{Special
bosonic dynamics is interpreted  in \cite{Ivanov+} 
by means of Eq. (5.2) there
as that describing electrically charged particle
in spherically symmetric electric
 and magnetic fields of the 
(non-(anti)self-dual) form  $\mathcal{E}_i=\nu^2 r_i/r^4$
and  $\mathcal{B}_i=-\nu r_i/r^3$, cf. \cite{CromRit,DHokVin3} 
and our discussion in Section 3.}.
We have shown that the superextension of the system,
which corresponds  to an (anti)self-dual dyon 
background, is described by
a qudratically extended Lie superalgebra $D(2,1;\alpha)$
with the parameter $\alpha=1/2$.  This belongs to the set of 
the values 
$\{-3,-3/2,-2/3,-1/3,1/2,2\}$, which can be related by
the superalgebra automorphism $D_3$, see footnote 3. 
The case of $\alpha=-1/2$, also
appearing in special bosonic 
dynamics in \cite{Ivanov+},
is not included in the indicated set.  As it is clear, particularly,
 from the form of the $D(2,1;\alpha)$ Casimir 
 (\ref{Casimir}), the case $\alpha=-1/2$ 
 (related by the $D_3$ automorphism to the cases
 $\alpha=-2$ and $\alpha=1$) is essentially different.
 For this value of the parameter $\alpha$ 
 the Casimirs of the two $so(3)$ subalgebras 
 enter  with the same ``weight" into the 
 $D(2,1;\alpha)$ Casimir. 
 Therefore, it seems to be very interesting 
 to investigate another special case $\alpha=-1/2$
 of the  $D(2,1;\alpha)$ model 
  \cite{Ivanov+}  from the viewpoint of nonlinear 
 extension of superconformal symmetry 
 associated with a presence in the bosonic 
 sector of the Laplace-Runge-Lenz vector and 
its dynamical integral counterpart. 
\vskip0.1cm

As it was already noted in the introduction, for one-dimensional 
quantum systems with soliton potentials the hidden free nature 
of the systems reveals itself in two related ways. Namely, such systems
are characterized by zero reflection coefficient, and they possess
the Lax-Novikov higher order  differential operator  as 
a nontrivial integral.  The latter is a Darboux-dressed form 
of the free particle momentum operator, and its differential 
order is fixed by the number of solitons `hidden' in the potential.
The hidden free nature also shows up in 
a supersymmetric extension of a reflectionless system.
Instead of the two supercharges  which we have in ordinary,
superextended systems with two  superpartner Hamiltonians, the 
solitonic superextended systems admit four
supercharges. Together  with the two bosonic integrals associated with 
the Lax-Novikov integrals, they generate a nontrivial nonlinear 
superalgeba \cite{AMP}. The  3D quantum system we studied here 
also reveals a hidden, partially free dynamics. 
The  system (\ref{Hspec}) is characterized  by a vanishing scattering 
angle for all values of $J$, $J^2\geq \nu^2$, and this property
is reminiscent of the vanishing reflection coefficient in 1D
solitonic systems. Thus it may be interesting to investigate whether 
the system we studied here possesses yet an additional hidden 
nonlinear supersymmetric structure associated with a partially free 
nature of its dynamics. This is suggested by the property of
the 1D conformal quantum mechanical model with $a/x^2$ potential.
It has a hidden, bosonized supersymmetric structure for special values 
of the strength $a$ \cite{fictsusy}.

\vskip0.2cm

\noindent \textbf{Acknowledgements.}
We thank  Sergey Fedoruk and Olaf Lechtenfeld  for helpful comments.
The work  has been partially supported by
FONDECYT Grant 1130017, and by  DICYT (USACH), 
and by DFG-Grants Wi 777/11 and GRK 1523.
MSP  and AW  are  grateful, respectively, 
to  the Universities of Jena and Santiago de Chile
for hospitality.



\begin{thebibliography}{99}

\bibitem{Pauli}
W. Pauli, \emph{``\"Uber das Wasserstoffspektrum vom Standpunkt der neuen Quantenmechanik,"} 
Z. Physik {\bf 36}, 336 (1926).

\bibitem{Fock}
V.~Fock,
\emph{  ``Zur Theorie des Wasserstoffatoms"},
Z.\ Phys.\  {\bf 98}, 145 (1935).


\bibitem{Barg}
V. Bargmann, \emph{  ``Zur Theorie des 
Wasserstoffatoms: Bemerkungen zur gleichnamigen 
Arbeit von V. Fock,"}  Z. Phys. {\bf 99}, 576 (1936).

\bibitem{Gold}
H. Goldstein, 
\emph{  ``Prehistory of the "Runge-Lenz" vector,"} 
Am. J. Phys. {\bf 43},  737 (1975); 
\emph{  ``More on the prehistory of the Laplace or 
RungeÐLenz vector,"}  Am. J. Phys. {\bf 44},  1123 (1975).

\bibitem{KLPW} 
 A.~Kirchberg, J.~D.~Lange, P.~A.~G.~Pisani and A.~Wipf,
\emph{  ``Algebraic solution of the supersymmetric hydrogen 
  atom in d-dimensions,''}
  Annals Phys.\  {\bf 303}, 359 (2003)
  [hep-th/0208228].
 
  
\bibitem{Wipf:2005se} 
  A.~Wipf, A.~Kirchberg and J.~D.~Lange,
  \emph{``Algebraic solution of the supersymmetric hydrogen atom,''}
  Bulg.\ J.\ Phys.\  {\bf 33}, 206 (2006)
  [hep-th/0511231].
 


\bibitem{NovZak}
S. P. Novikov, S.V. Manakov, L. P. Pitaevskii, and V. E. Zakharov,
\emph{Theory of Solitons} (Plenum, New York, 1984).

\bibitem{AMP} 
  A.~Arancibia, J.~M.~Guilarte and M.~S.~Plyushchay,
\emph{  ``Effect of scalings and translations on the 
  supersymmetric quantum mechanical structure of soliton systems,''}
  Phys.\ Rev.\ D {\bf 87}, 045009 (2013)
  [arXiv:1210.3666 [math-ph]];
\emph{  ``Fermion in a multi-kink-antikink soliton background, 
  and exotic supersymmetry,''}
  Phys.\ Rev.\ D {\bf 88}, 085034 (2013)
  [arXiv:1309.1816 [hep-th]].

\bibitem{GoddOl} 
  P.~Goddard and D.~I.~Olive,
 \emph{ ``Magnetic monopoles in gauge field theories,''}
  Rept.\ Prog.\ Phys.\  {\bf 41}, 1357 (1978).

\bibitem{PlMon} 
  M.~S.~Plyushchay,
  \emph{``Monopole Chern-Simons term: 
  Charge monopole system as a particle with spin,''}
  Nucl.\ Phys.\ B {\bf 589}, 413 (2000)
  [hep-th/0004032];
 \emph{ ``Free conical dynamics: 
  Charge-monopole as a particle with spin, 
  anyon and nonlinear fermion-monopole supersymmetry,''}
  Nucl.\ Phys.\ Proc.\ Suppl.\  {\bf 102}, 248 (2001)
  [hep-th/0103040].
 

\bibitem{Zwanz} 
  D.~Zwanziger,
  \emph{ ``Exactly soluble nonrelativistic model 
  of particles with both electric and magnetic charges,''}
  Phys.\ Rev.\  {\bf 176}, 1480 (1968).
 


\bibitem{Mcintosh:1970gg} 
  H.~V.~Mcintosh and A.~Cisneros,
\emph{  ``Degeneracy in the presence of a magnetic monopole,''}
  J.\ Math.\ Phys.\  {\bf 11}, 896 (1970).


\bibitem{Witten} 
  E.~Witten,
 \emph{ ``Dyons of charge $e \theta/2 \pi$,''}
  Phys.\ Lett.\ B {\bf 86}, 283 (1979).

\bibitem{Jackiw} 
  R.~Jackiw,
\emph{  ``Dynamical symmetry of the magnetic monopole,''}
  Annals Phys.\  {\bf 129}, 183 (1980).
 

\bibitem{DHokVin1} 
  E.~D'Hoker and L.~Vinet,
 \emph{ ``Constants of motion 
 for a spin 1/2 particle in the field of a dyon,''}
  Phys.\ Rev.\ Lett.\  {\bf 55}, 1043 (1985).

\bibitem{DHokVin2} 
  E.~D'Hoker and L.~Vinet,
 \emph{ ``Hidden symmetries and accidental degeneracy 
  for a spin 1/2 particle in the field of a dyon,''}
  Lett.\ Math.\ Phys.\  {\bf 12}, 71 (1986).

\bibitem{DHokVin3}
  E.~D'Hoker and L.~Vinet,
\emph{ ``Dynamical supersymmetry of the 
magnetic monopole and the $1/r^2$ potential,''}
  Commun.\ Math.\ Phys.\  {\bf 97} (1985) 391.


\bibitem{HorP} 
 L.~G.~Feher and P.~A.~Horvathy,
 \emph{  ``Non-relativistic scattering of a spin-1/2 particle off a self-dual monopole,''}
  Mod.\ Phys.\ Lett.\ A {\bf 3}, 1451 (1988)
  [arXiv:0903.0249 [hep-th]];
  L.~Feher, P.~A.~Horvathy and L.~O'Raifeartaigh,
\emph{  ``Separating the dyon system,''}
  Phys.\ Rev.\ D {\bf 40}, 666 (1989);
\emph{  ``Applications of chiral supersymmetry for 
  spin fields in selfdual backgrounds,''}
  Int.\ J.\ Mod.\ Phys.\ A {\bf 4}, 5277 (1989)
  [arXiv:0903.2920 [hep-th]];
  F.~Bloore and P.~A.~Horvathy,
\emph{  ``Helicity supersymmetry of dyons,''}
  J.\ Math.\ Phys.\  {\bf 33}, 1869 (1992)
  [hep-th/0512144].


\bibitem{DeJMPH} 
  G.~W.~Gibbons, R.~H.~Rietdijk and J.~W.~van Holten,
 \emph{ ``SUSY in the sky,''}
  Nucl.\ Phys.\ B {\bf 404}, 42 (1993)
  [hep-th/9303112];
  F.~De Jonghe, A.~J.~Macfarlane, K.~Peeters and J.~W.~van Holten,
\emph{  ``New supersymmetry of the monopole,''}
  Phys.\ Lett.\ B {\bf 359}, 114 (1995)
  [hep-th/9507046].


\bibitem{PnSU} 
  M.~S.~Plyushchay,
 \emph{ ``On the nature of fermion-monopole supersymmetry,''}
  Phys.\ Lett.\ B {\bf 485}, 187 (2000)
  [hep-th/0005122].
  

\bibitem{Horvathy:2000cc} 
  P.~A.~Horvathy, A.~J.~Macfarlane and J.~W.~van Holten,
 \emph{ ``Monopole supersymmetries 
 and the Biedenharn operator,''}
  Phys.\ Lett.\ B {\bf 486}, 346 (2000)
  [hep-th/0006118].


\bibitem{Papadopoulos:2000ka} 
  G.~Papadopoulos,
 \emph{ ``Conformal and superconformal mechanics,''}
  Class.\ Quant.\ Grav.\  {\bf 17}, 3715 (2000)
  [hep-th/0002007].


\bibitem{de Azcarraga:2001mg} 
  J.~A.~de Azcarraga, J.~M.~Izquierdo and A.~J.~Macfarlane,
\emph{  ``Hidden supersymmetries in supersymmetric quantum mechanics,''}
  Nucl.\ Phys.\ B {\bf 604}, 75 (2001)
  [hep-th/0101053].
 

\bibitem{Cotaescu:2001re} 
  I.~I.~Cotaescu and M.~Visinescu,
 \emph{ ``Runge-Lenz operator for Dirac field in Taub - NUT background,''}
  Phys.\ Lett.\ B {\bf 502}, 229 (2001)
  [hep-th/0101163].
 

\bibitem{Ivanov+} 
  E.~Ivanov, S.~Krivonos and O.~Lechtenfeld,
\emph{  ``New variant of N=4 superconformal mechanics,''}
  JHEP {\bf 0303}, 014 (2003)
  [hep-th/0212303].

\bibitem{IvaLech}
  E.~Ivanov and O.~Lechtenfeld,
\emph{    ``N=4 supersymmetric mechanics in harmonic superspace,''}
  JHEP {\bf 0309}, 073 (2003)
  [hep-th/0307111].


\bibitem{LePl} 
  C.~Leiva and M.~S.~Plyushchay,
 \emph{ ``Nonlinear superconformal symmetry 
  of a fermion in the field of a Dirac monopole,''}
  Phys.\ Lett.\ B {\bf 582}, 135 (2004)
  [hep-th/0311150].


\bibitem{Bellucci:2006gs} 
  S.~Bellucci, A.~Nersessian and A.~Yeranyan,
\emph{  ``Hamiltonian reduction and 
  supersymmetric mechanics with Dirac monopole,''}
  Phys.\ Rev.\ D {\bf 74}, 065022 (2006)
  [hep-th/0606152].
  

\bibitem{AveMich} 
  S.~G.~Avery and J.~Michelson,
 \emph{ ``Mechanics and quantum supermechanics of a 
 monopole probe 
  including a Coulomb potential,''}
  Phys.\ Rev.\ D {\bf 77}, 085001 (2008)
  [arXiv:0712.0341 [hep-th]].
 

\bibitem{Ngome:2010gg} 
  J.~P.~Ngome, P.~A.~Horvathy and J.~W.~van Holten,
\emph{  ``Dynamical supersymmetry of spin 
  particle-magnetic field interaction,''}
  J.\ Phys.\ A {\bf 43}, 285401 (2010)
  [arXiv:1003.0137 [hep-th]].


\bibitem{ADKS} 
 D.~Anninos, T.~Anous, F.~Denef, G.~Konstantinidis and E.~Shaghoulian,
 \emph{ ``Supergoop dynamics,''}
  JHEP {\bf 1303}, 081 (2013)
  [arXiv:1205.1060 [hep-th]].

\bibitem{KNO} 
  S.~Krivonos, A.~Nersessian and V.~Ohanyan,
  \emph{``Multi-center MICZ-Kepler system, 
  supersymmetry and integrability,''}
  Phys.\ Rev.\ D {\bf 75}, 085002 (2007)
  [hep-th/0611268].


\bibitem{Hong} 
  S.~-T.~Hong, J.~Lee, T.~H.~Lee and P.~Oh,
\emph{  ``A complete solution of a constrained system: 
  SUSY monopole quantum mechanics,''}
  JHEP {\bf 0602}, 036 (2006)
  [hep-th/0511275].

  
\bibitem{conf1} 
  V.~de Alfaro, S.~Fubini and G.~Furlan,
\emph{  ``Conformal invariance in quantum mechanics,''}
  Nuovo Cim.\ A {\bf 34}, 569 (1976).

\bibitem{confla} 
 J.~Michelson and A.~Strominger,
 \emph{ ``The geometry of (super)conformal quantum mechanics,''}
  Commun.\ Math.\ Phys.\  {\bf 213}, 1 (2000)
  [hep-th/9907191].

\bibitem{Claus:1998ts} 
  P.~Claus, M.~Derix, R.~Kallosh, J.~Kumar, 
  P.~K.~Townsend and A.~Van Proeyen,
\emph{  ``Black holes and superconformal mechanics,''}
  Phys.\ Rev.\ Lett.\  {\bf 81}, 4553 (1998)
  [hep-th/9804177].
 
\bibitem{LeiPlyAdS} 
  C.~Leiva and M.~S.~Plyushchay,
  \emph{``Conformal symmetry of relativistic and 
  nonrelativistic systems and AdS/CFT correspondence,''}
  Annals Phys.\  {\bf 307}, 372 (2003)
  [hep-th/0301244].


\bibitem{Fedoruk:2009pq} 
  S.~Fedoruk, E.~Ivanov and O.~Lechtenfeld,
\emph{  ``OSp(4$|$2) superconformal mechanics,''}
  JHEP {\bf 0908}, 081 (2009)
  [arXiv:0905.4951 [hep-th]].


  \bibitem{confla+} 
  S.~Fedoruk, E.~Ivanov and O.~Lechtenfeld,
\emph{  ``New D(2,1;$\alpha$) mechanics with spin variables,''}
  JHEP {\bf 1004}, 129 (2010)
  [arXiv:0912.3508 [hep-th]].


\bibitem{CromRit}
 M.~de Crombrugghe and V.~Rittenberg,
 \emph{  ``Supersymmetric quantum mechanics,''}
  Annals Phys.\  {\bf 151}, 99 (1983).


\bibitem{VDJ} 
  J.~Van Der Jeugt,
 \emph{ ``Irreducible representations of the 
  exceptional Lie superalgebras D(2,1;$\alpha$),''}
  J.\ Math.\ Phys.\  {\bf 26}, 913 (1985).

\bibitem{FSS} 
  L.~Frappat, P.~Sorba and A.~Sciarrino,
 \emph{ ``Dictionary on Lie superalgebras,''}
  hep-th/9607161.
 

\bibitem{BMSV} 
  R.~Britto-Pacumio, J.~Michelson, A.~Strominger and A.~Volovich,
 \emph{ ``Lectures on superconformal quantum 
  mechanics and multiblack hole moduli spaces,''}
  hep-th/9911066.

\bibitem{IKL} 
  E.~Ivanov, S.~Krivonos and O.~Lechtenfeld,
  \emph{``N=4, d = 1 supermultiplets from 
  nonlinear realizations of D(2,1; $\alpha$),''}
  Class.\ Quant.\ Grav.\  {\bf 21}, 1031 (2004)
  [hep-th/0310299].

\bibitem{MatMor} 
  T.~Matsumoto and S.~Moriyama,
 \emph{ ``An exceptional algebraic origin of the AdS/CFT 
  Yangian symmetry,''}
  JHEP {\bf 0804}, 022 (2008)
  [arXiv:0803.1212 [hep-th]].

\bibitem{FIOL} 
  S.~Fedoruk, E.~Ivanov and O.~Lechtenfeld,
\emph{  ``Superconformal mechanics,''}
  J.\ Phys.\ A {\bf 45}, 173001 (2012)
  [arXiv:1112.1947 [hep-th]].


\bibitem{Dirac} 
  P.~A.~M.~Dirac,
 \emph{``Quantized singularities in the electromagnetic field,''}
  Proc.\ Roy.\ Soc.\ Lond.\ A {\bf 133}, 60 (1931).

\bibitem{KirLanWip} 
  A.~Kirchberg, J.~D.~Lange and A.~Wipf,
  \emph{``Extended supersymmetries and the Dirac operator,''}
  Annals Phys.\  {\bf 315}, 467 (2005)
  [hep-th/0401134].

  
\bibitem{gamma5} 
  M.~S.~Plyushchay,
 \emph{``Pseudoclassical description 
 of the massive spinning particle in d-dimensions,''}
  Mod.\ Phys.\ Lett.\ A {\bf 8}, 937 (1993);
    J.~Gamboa and M.~Plyushchay,
\emph{  ``Classical anomalies for spinning particles,''}
  Nucl.\ Phys.\ B {\bf 512}, 485 (1998)
  [hep-th/9711170].
 

 \bibitem{Zwanz2}
 D. Zwanziger, 
 \emph{  ``Algebraic calculation of nonrelativistic Coulomb phase shifts,"},
 J. Math. Phys. {\bf 8}, 1858 (1967). 

\bibitem{fictsusy} 
  F.~Correa, M.~A.~del Olmo and M.~S.~Plyushchay,
\emph{  ``On Hidden broken nonlinear superconformal 
  symmetry of conformal mechanics and nature 
  of double nonlinear superconformal symmetry,''}
  Phys.\ Lett.\ B {\bf 628}, 157 (2005)
  [hep-th/0508223].
  
\end{thebibliography}
\end{document}